\documentclass[aps,prb,floats,12pt]{revtex4}
\input{epsf}
\input{psfig.sty}
\usepackage{amsmath}
\usepackage{amssymb}
\usepackage{graphicx}
\begin{document}
\title{Mechanisms of cooperativity underlying sequence-independent
$\beta$-sheet formation}
\author{Chinlin Guo, Margaret S. Cheung, Herbert Levine}
\affiliation{Department of Physics, University of California, San Diego\\
9500 Gilman Drive, La Jolla, CA 92093-0319}
\author{David A. Kessler}
\affiliation{Department of Physics, Bar-Ilan University, Ramat-Gan, Israel}
\date{\today}
\def\ra{{r_{\rm Hb}^c}}
\def\rb{{r_{\rm Hb}^*}}
\def\rc{{r_{\rm Hb}^{sol}}}
\def\rx{{r_{\rm Hb}^x}}
\def\x{{x_{\rm Hb}}}
\def\Ra{{r_{\rm sc}^c}}
\def\Rb{{r_{\rm sc}^*}}
\def\Rc{{r_{\rm sc}^{sol}}}
\def\Rx{{r_{\rm sc}^x}}
\def\X{{x_{\rm chiral}}}
\def\dt{{\Delta t}}
\def\go{{G$\rm\bar{o}$}}
\def\rHb{{NH$\cdots$OC}}
\maketitle
\begin{center} {\bf ABSTRACT} \end{center} 

{\bf
We investigate the formation of $\beta$-sheet structures in
proteins without sequence-dependent side-chain interactions.
To accomplish this, we introduce a model
which explicitly incorporates both solvation effects and the
angular dependence (on the protein backbone) of hydrogen bond
formation. The thermodynamics of this model is studied by
exploring the density of states for the entire system and
the local couplings in a partially folded structure.
Our results suggest that solvation dynamics together with the
H-bond angular dependence gives rise to a generic cooperativity
in this class of systems; this result explains why pathological aggregates
involving $\beta$-sheet cores can form from many different proteins. Our
work provides the foundation for the construction of
phenomenological models to investigate topology effects in $\beta$-sheet
folding and the competition between native folding and non-specific
aggregation.
}

\section{Introduction} 

$\beta$-sheets are an important element of protein structure,
occurring often both in functional units and in pathological
aggregates. For example, signaling proteins such as SH3 contain a
$\beta$ core. On the other hand, the precursors of amyloidogenesis
are commonly found to have a $\beta$-sheet organization
\cite{Sunde}; moreover, a conformational rearrangement can switch
a functional $\beta$-sheet into an aggregation unit
\cite{Kelly1,Kelly2}. As they play the role of stabilizing highly
organized architectures (either pathological or functional),
$\beta$-sheets must be strongly cooperative units, responsible for
lowering the free energy of these folded structures as compared to
entropy-dominated more irregular patterns.

It is often assumed that hydrophobic interactions are the major forces
contributing to $\beta$-sheet cooperativity and their interacting
pattern (i.e., the sequence design) can define a unique fold \cite{Fersht}.
But, one cannot ignore the
fact that many neurodegenerative diseases are caused by the
aggregation of mutant proteins with long {\em hydrophilic}
sequences (e.g., poly-glutamine in Huntington disease
\cite{Perutz}); the $\beta$-sheet at the core of these aggregates
are certainly not being stabilized by hydrophobic forces. Indeed,
recent experiments have shown that peptides with purely
hydrophilic sequences can fold $\beta$-sheets cooperatively
\cite{Engelman}. Also, growing evidence suggests that proteins are
capable of forming $\beta$-sheet aggregates, regardless of their
native folds \cite{Dobson}. Together with the fact that most
amyloidogenetic precursors do not share any homologous sequence
\cite{Sunde}, it is therefore very likely that throughout the
2-dimensional $\beta$-sheet architecture, formation of an H-bond
network can generate long-range ordering, hence compensating the
mismatch of side-chain interactions and stabilizing the system.

What are the basic differences between backbone H-bond and
side-chain hydrophobic interactions? Besides the difference in
sequence specificity discussed above, they also differ in the
maximal number of contacts for each interaction unit. Along the
peptide chain, one amino acid can form at most
two backbone H-bonds with adjacent $\beta$-strands in the
2-dimensional $\beta$-sheet structure. In contrast, for each hydrophobic
residue in a compact protein, the number of
favorable (sequence specific) contacts can be more than two; this
leads to many-body effects that can define and stabilize a unique fold
\cite{Fersht,Serrano,Dokholyan}. In most protein folding literature,
the hydrophobic many-body effect is mimicked by the use of
\go-type interaction \cite{Go} and sequence-specific restrictions on
the backbone energy \cite{cecilia} to favor the native state. This
approach, however, is not generally applicable to the case of
sequence-independent $\beta$-sheet formation.

Moreover, from a statistical mechanics prospective, it is well known
that having cooperative behavior in a 2-dimensional,
pairwise-interacting system requires more than two couplings (in
average) for each interaction unit \cite{SM}. In protein folding,
this corresponds to the case that folding of a specific part of
the overall structure will aid its contiguous contact formation
without redundant entropy loss, as briefly discussed by Berriz {\it et al}
\cite{Berriz}. At the first glance, this principle does not apply
to the case of H-bond network formation in an obvious way. Thus,
elucidating how the networking of individual H-bond can give rise
to a stable folded structure is the major goal of present work.

For this purpose, we study a model system without specific
side-chain interaction. It has already been shown that for
$\beta$-structured proteins with randomly assigned sequences, a
pure Lennard-Jones potential might not be sufficient to provide
systemic cooperativity, \cite{Brooks} and angular dependence in
contact formation must be taken into account \cite{Berriz}. Also,
there is growing evidence that solvent effects play a significant
role in protein folding and conformational changes
\cite{Hummer,water,Dunitz,tension,VWF1}. Thus, we introduce a new
approach to integrate these two effects in modeling H-bond
formation. The ingredients of our approach are a ``co-plane''
parameterization of the backbone degrees of freedom with explicit
couplings of solvation dynamics and the angular dependence of
hydrogen-bond formation. The systemic cooperativity is
investigated by exploring the density of states for $\beta$-sheets
of different size and structure. To facilitate the sampling, we
have used replica-exchange \cite{replica} and multi-canonical
rescaling \cite{MC} techniques. The origin of this global
cooperativity is further studied by comparing energetics and
thermodynamic stability of local sub-structures of model systems
with different types of interaction potentials.

Our results indicate that it is the de-solvation cavity formation (i.e.,
the granularity of the water molecule) and the hydrogen-bond angular dependence
that provide $\beta$-sheets with a systemic cooperativity.
Since this property is sequence-independent, our results suggest why
$\beta$-sheets can generically serve as the building blocks for constructing
functional macromolecules, and why most functional proteins have the tendency
to form aggregates at sufficiently high concentration.

\section{Method}

\subsection{The model system}

The system proposed here is based on the ``ball-and-stick'' model introduced
in Ref. \onlinecite{Peter}. In this approach, all NH, C$_\alpha$, CO
(represented as $\rm C'$), and C$_\beta$ side-chains are treated
as lumped balls. The local backbone energy consists of a sum
of terms:
\begin{enumerate}
\item bond bending: ${\rm V}_\theta=\frac{1}{2}\sum
k_{\theta}(\theta_i-\theta_0)^2$
\item dihedral rotations: ${\rm
V}_\omega=\frac{1}{2}\sum \epsilon_{\omega}(1+\cos\omega_i)$, ${\rm
V}_\phi=\frac{1}{2}\sum \epsilon_{\phi}(1+\cos 3\phi_i)$, ${\rm
V}_\psi=\frac{1}{2}\sum \epsilon_{\psi}(1+\cos 3\psi_i)$,
\item side-chain chirality: ${\rm V}_{chiral}=\sum\epsilon_{chiral}
\Theta\left(\left\{[\hat{r}_{NC_\alpha}\times\hat{r}_{C_\beta
C_\alpha}]\cdot \hat{r}_{C'C_\alpha}\right\}_i\right)$
\end{enumerate}
with the sum taken
over all components (labeled by $i$), and $\Theta(s)=(s+|s|) /2$. Here,
since we are interested in sequence-independent effects, all
side-chains are treated as polar or weakly charged groups that
favor $\beta$-sheet formation, with interaction strengths
\cite{Fersht,Protein} that are far less than the magnitude of the H-bond
energy and can be generally ignored. Also, we take the parameters
$k_{\theta}=200 \rm kcal/mol\cdot rad^2$, $\epsilon_{\omega}=40
\rm kcal/mol$, $\epsilon_{\phi}=\epsilon_{\psi}=0.45 \rm
kcal/mol$, and $\epsilon_{chiral}=10 \rm kcal/mol$ from
Ref. \onlinecite{Peter}.

\subsection{The solvation force field} 

Next, we construct expressions for the non-local interactions.
We incorporate the solvation effect into the force law
by designing a double-well potential that can
account for contact formation and single-H$_2$O hydration (between
the contact pair); the barrier between the wells corresponds to
the free energy cost involved in the transfer of a water molecule
out of the hydration shell \cite{Hummer,Helson}. Multiple shells
could be accommodated via a multiple-well potential, but this is not attempted
here. The approximation of using an effective free-energy is justified via
the observation that transfer of water from within the vicinity
of the contact pair is faster than the contact formation
\cite{Hummer,water,Dunitz,tension}.

There is no standard way to construct a continuous de/solvation
potential \cite{solvent} which can be used for Langevin
simulations; we note that a previous curve fitting effort \cite{Stillinger}
and a recent Monte Carlo study
\cite{Helson} utilized a discrete version of such a potential. We
therefore propose an empirical profile based on estimated
experimental parameters. Specifically, we will take for H-bond
formation
\begin{eqnarray}\label{Solv_Go}
{\rm U_{Hb}}(r)&=&\frac{A}{r^k}\Biggl[\frac{1}{4r^{3k}}
-\left(\frac{1}{(\ra)^k}+\frac{1}{(\rb)^k}+\frac{1}{(\rc)^k}\right)
\frac{1}{3r^{2k}}\nonumber\\
&&\hspace{0.5cm}
+\left[\left(\frac{1}{(\ra)^k}+\frac{1}{(\rb)^k}\right)\frac{1}{(\rc)^k}
+\frac{1}{(\ra\rb)^k}\right]\frac{1}{2r^{k}}
-\frac{1}{(\ra\rb\rc)^k}\Biggr]
\end{eqnarray}
Here $r$ is the distance between NH and CO, and the integer $k\in
N$ ($k=6$, e.g.) is chosen to give specific long-range behavior.
The values $r=\ra$, $r=\rb$, and $r=\rc$ represent the separation
distance at the contact bond position, at the the peak of the desolvation
barrier, and in the presence of a single
intervening solvent molecule separation, respectively.
$\rc$ is obtained by adding the known value of $\ra$ to the size
of a single H$_2$O. $A$ and $\rb$ can be determined by the
strengths of the H-bond \cite{Peter} and the desolvation barrier;
the latter is approximated by the surface energy involved in
forming a desolvation cavity, estimated to equal $\rm\approx 0.103
kcal/(mol\stackrel{\circ}{A^2})$ \cite{tension}. Here we have
$\ra\approx 3.43\stackrel{\circ}{\rm A}$ and
$\rb\approx 3.92\stackrel{\circ}{\rm A}$; these yield a single
H$_2$O hydration energy of $\approx -0.5$ kcal/mol.

Clearly, the design principle behind this type of formula is to have
three roots for $d {\rm U_{Hb}}(r)/dr=0$ accounting for the
desolvation barrier and for the two local minima; one can add more
roots to address multiple solvation shell effects. Note that one
might wish to fine-tune the potential profile more precisely
(for instance, by being able to independently vary the width of
the contact well). This can be accomplished by a more complicated
expression (see appendix \ref{appen}). Also, the strength of
de/solvation might depend on the hydrophobicity of local environment
(i.e., de-wetting behavior) and one can certainly take this into account.

As discussed in \cite{Peter,Protein}, hydrogen bond formation has
a strong angular dependence on its surrounding backbone; thus, one
can not fully describe the interaction using the radial distance
alone. From Fig. \ref{fig0}(d), we note that such an angular
dependence is merely the requirement that all atoms near to the
interacting NH, CO (i.e. the shadowed area) are aligned
``natively". In Ref. \onlinecite{Peter}, this ``alignment'' effect
is accomplished by introducing artificial repulsive forces between
H-bond neighbors. Here, we propose instead a ``functional-block''
(co-plane) scheme.

\subsection{The co-plane approach} 

From Fig. \ref{fig0}(a), we notice that the peptide backbone
between two contiguous C$_\alpha$ atoms has a co-planar
structure because of the N-C-O bond resonance. This allows us to
model the motion of a ``co-plane'' as a whole (i.e., with one
dihedral angle $\omega$ fixed to $\approx 180^{\circ}$). Overall,
each interacting unit in our system is just one block of C$_\alpha$
co-plane. The
$i^{\rm th}$ co-plane is defined by three points $\{\rm
X_1^{(i)},X_2^{(i)},X_3^{(i)}\}$ where $\rm X_1^{(i)},X_3^{(i)}$
are simply the two C$_\alpha$ in the plane and $\rm X_2^{(i)}$ is
a virtual point satisfying $\overrightarrow{\rm
X_1^{(i)}X_2^{(i)}}\perp \overrightarrow{\rm X_2^{(i)}X_3^{(i)}}$,
fig.\ref{fig0}(b). Once the degree of freedom corresponding to
this point is specified,  all the atomic locations of the amide
and carbonyl groups are fixed. Note that a convenient way to
define the arbitrary point $\rm X_2$ is such that the vector
$\overrightarrow{\rm X_2^{(i)}X_3^{(i)}}$ points directly at the
C$_\alpha$ of the neighboring strand if the system is in the
native $\beta$-sheet structure; see fig.\ref{fig0}(c). Once ${\rm
X}_2^{(i)}$ is chosen, this determines the in-plane angles
$\theta_1$ and $\theta _2$ as well as the virtual plane projection
angle $\theta_3$. We computed all these objects by using the
native value of bond angles \cite{Peter} and dihedral angles
observed in regular $\beta$-sheets ($\phi\sim-138\pm1^{\circ}$,
$\psi\sim135\pm 1^{\circ}$) \cite{Protein}. This leads to all
geometrical properties of the virtual plane, including
$\theta_1=64.7^{\circ}$, $\theta_2=62.2^{\circ}$,
$\theta_3=22.4^{\circ}$,
$\rm\overline{X_1X_2}=3.75\stackrel{\circ}{\rm A}$,
$\rm\overline{X_2X_3}=0.57\stackrel{\circ}{\rm A}$.

Given the native plane, we define the range of possible non-native
structures via allowing the orientation of the planes to vary.
This yields three degrees of freedom per residue; two reflect the
angles needed to define the direction of the fixed length vector
going from one $C_\alpha$ to the next while the third refers to a
rolling of the plane around this vector. One can then work out the
geometrical problem of expressing, in terms of these degrees of
freedom, the residual terms in the backbone energy arising from
the dihedral terms involving $\phi$ and $\psi$, the bending of the
bond angle $\rm \angle C' C_\alpha N$, and the side-chain
chiralities between consecutive blocks. Our approach fully maintains the
overall translation and rotational degrees of freedom of the protein molecule.
This co-plane parameterization
greatly reduces our computation effort as compared to all atom
backbones, yet maintains the roll degree of freedom not present in
the simplified C$_\alpha$ model. As will be shown later, this
degree of freedom is important for the study of $\beta$-sheet
cooperativity.

\subsection{The structural factor}

Now we use the co-plane approach to model the H-bond angular dependence.
From fig.\ref{fig0}(e), we note that having a H-bond between blocks
$i,j$ requires fixing two rotational and two translational degrees
of freedom. More precisely, we need $\overrightarrow{\rm
X_1^{(i)}X_2^{(i)}}$ antiparallel to $\overrightarrow{\rm
X_1^{(j)}X_2^{(j)}}$, and $\overrightarrow{\rm
X_2^{(i)}X_3^{(i)}}$ parallel to $\overrightarrow{\rm
X_2^{(j)}X_3^{(j)}}$, whereas for translational alignment, the two
sets of points  $\{\rm X_1^{(i)},X_2^{(j)},X_3^{(j)}\}$ and $\{\rm
X_1^{(j)},X_3^{(i)},X_2^{(i)}\}$ must be collinear. This leads us
to define a structural factor that monitors the angular
nativeness in the H-bond
\begin{eqnarray}\label{xHb}
x_{\rm Hb}^{(i,j)}
&=&
=
\frac{1}{16}\rm
\left[1-\widehat{X_{12}^{(i)}}\cdot\widehat{X_{12}^{(j)}}\right]
\left[1+\widehat{X_{23}^{(i)}}\cdot\widehat{X_{23}^{(j)}}\right]
\nonumber\\&&\hspace{0.5cm}\times\rm
\left[1+\cos \angle X_2^{(j)}X_1^{(i)} X_3^{(j)}\right]
\left[1+\cos \angle X_2^{(i)}X_1^{(j)} X_3^{(i)}\right]
\end{eqnarray}
where $\rm\widehat{X_{ab}^{(k)}}$ is the unit vector connecting
$\rm X_a^{(k)},X_b^{(k)}$. Note that $x_{\rm Hb}^{(i,j)}$ has maximum of 1
only when all the alignment criteria are satisfied.

We now wish to incorporate this structural factor into the Hamiltonian
in such a manner as to ensure that at small $x_{\rm Hb}$, the two
interacting blocks are unlikely to form H-bond. A simple
phenomenological way to proceed is to introduce a potential
profile which has two parameters dynamically modulated by $x_{\rm Hb}$ with a
fixed solvation energy $E_{sol}$
\begin{eqnarray}\label{Solv_ang}
&&{\rm U_{Hb}}(r,\x)=
\frac{A(\x)}{r^k}\Biggl[\frac{1}{4r^{3k}}
-\left(\frac{1}{\rx^k}+\frac{1}{(\rb)^k}+\frac{1}{(\rc)^k}\right)
\frac{1}{3r^{2k}}\nonumber\\
&&\hspace{1.0cm}
+\left[\left(\frac{1}{\rx^k}+\frac{1}{(\rb)^k}\right)\frac{1}{(\rc)^k}
+\frac{1}{(\rx\rb)^k}\right]\frac{1}{2r^{k}}
-\frac{1}{(\rx\rb\rc)^k}\Biggr]
\end{eqnarray}
Here the first dynamical parameter $A(\x)$ is determined by fixing ${\rm
U_{Hb}}(\rc,\x)=E_{sol}$ and for another parameter, we have used a linear
relation $\rx=\x\ra+(1-\x)\rb$. Examples of this modulated double-well
profile are shown in fig.\ref{fig1}(a). Note that even incorporated with
this structural factor, the H-bond potential still remains pairwise between
two interacting blocks.

Aside from the NH$\cdots$OC hydrogen bond, we include in our model the bonding
between C$_\alpha$-H$\cdots$O=C \cite{chemical}. This interaction has a
strong chirality \cite{Peter} and an angular dependence that
involves at least two contiguous blocks. From fig.\ref{fig0}(d),
we note that the native configuration for interacting, consecutive
blocks $i,i+1$ and $j,j+1$ (here $i(j)+1$ is the block next to
$i(j)$ along the same strand) requires $\widehat{\rm
X_{23}^{(i)}}$ ($\widehat{\rm X_{23}^{(j)}}$) antiparallel to
$\widehat{\rm X_{23}^{(i+1)}}$ ($\widehat{\rm X_{23}^{(j+1)}}$).
Therefore, in analogy to the aforementioned H-bond ,we propose a
structural factor for this additional interaction
\begin{eqnarray}\label{xsc}
x_{\rm chiral}^{(i,j)}=\frac{1}{4}\x^{(i,j)}\x^{(i+1,j+1)}
\rm
\left[1-\widehat{\rm X_{23}^{(i)}}\cdot\widehat{\rm X_{23}^{(i+1)}}\right]
\left[1-\widehat{\rm X_{23}^{(j)}}\cdot\widehat{\rm X_{23}^{(j+1)}}\right]
\end{eqnarray}
with an interaction potential ${\rm U_{C_\alpha H-OC}}(r,\X)$ taken to be
similar to ${\rm U_{Hb}}(r,\x)$.
We use a C$_\alpha$-H$\cdots$O=C interaction strength half that of a
single H-bond \cite{chemical}.

\subsection{The role of the force field}

To characterize the roles of structural factors and solvation effects
in $\beta$-sheet formation, we have studied four different forms of the
H-bond potential. These are:
\begin{itemize}
\item {\bf (A) LJ$^{fix}$:} ${\rm U}_{\rm LJ}^{fix}(r) =|E_{\rm
Hb}|\frac{(\ra)^6}{r^6}\left[\frac{(\ra)^6}{r^6}-2\right]$,
a L-J potential without angular dependence.
\item{\bf (B) Sol$^{fix}$:}
${\rm U}_{\rm Sol}^{fix}(r) \Rightarrow$
eqn.(\ref{Solv_Go}), a double-well solvation potential without
angular dependence.
\item {\bf (C) LJ$^{ang}$:} ${\rm U}_{\rm LJ}^{ang}(r) = A_{\rm LJ}(\x)
\frac{[r_{\rm LJ}(\x)]^6}{r^6}\left[\frac{[r_{\rm
LJ}(\x)]^6}{r^6}-2\right]$, a 2-dynamical-parameter L-J potential with our
proposed angular dependence. This profile is designed to help distinguish
the effects of solvation and structural factors. Thus, except for
the absence of a desolvation barrier, this potential is quite
similar to the solvation one, eqn.(\ref{Solv_ang}). Explicitly, we
let a non-native alignment factor $\x<1$ give rise to a linear
shift $|{\rm U}_{\rm LJ}^{ang}(r_{\rm LJ}(\x))| =A_{\rm
LJ}(\x)=\x|E_{\rm Hb}|+(1-\x)|E^{sol}|$ for the first parameter.
Also, to achieve a repulsive effect similar to that in eqn.(\ref{Solv_ang}),
we set ${\rm U}_{\rm LJ}^{ang}(\rx)={\rm U_{Hb}}(\rx,\x)$, i.e.,
$r_{\rm LJ}(\x)=\rx(1+\sqrt{1+{\rm U_{Hb}}(\rx,\x)/A_{\rm LJ}(\x)})^{1/6}$
for the second parameter. Fig.\ref{fig1}(b) shows the modulation of this
force field due to varying of structural factor.
\item {\bf (D) Sol$^{ang}$:} ${\rm U}_{\rm Solv}^{ang}(r)\Rightarrow$
eqn.(\ref{Solv_ang}), the de/solvation potential with the proposed
angular dependence, which constitutes our full model.
\end{itemize}

\subsection{The replica-exchange and Multi-canonical Technique}

Throughout the entire paper, the simulation procedure employed
a modified Verlet-Langevin algorithm. Specifically, for a particular set
of coordinate $\vec{x}(t)$ with velocity $\vec{v}(t)$ and force $\vec{f}(t)$,
the update at time $t$,
$\Delta\vec{x}(t)\equiv\vec{x}(t+\dt)-\vec{x}(t)$ is obtained by
\begin{eqnarray}\label{Verlet-Langevin}
\vec{x}(t\pm\dt)&=&\vec{x}(t)\pm\vec{v}\dt+\frac{\dt^2}{2m}\vec{f}+O(\dt^3)
\nonumber\\
\vec{x}(t+\dt)-\vec{x}(t-\dt)&=&2\vec{v}(t)\dt+O(\dt^3)\nonumber\\
\vec{x}(t+\dt)+\vec{x}(t-\dt)&=&2\vec{x}(t)+\frac{\dt^2}{m}\vec{f}(t)
+O(\dt^4)\nonumber\\
&=&2\vec{x}(t)+\frac{\dt^2}{m}\left[-\zeta\vec{v}(t)
-\vec{\nabla_x}E+\eta(t)\right]+O(\dt^4)\nonumber\\
\Rightarrow\Delta\vec{x}(t)&=&
\left[\frac{1-\frac{\zeta\dt}{2m}}{1+\frac{\zeta\dt}{2m}}\right]
\Delta\vec{x}(t-\dt)
+\frac{\dt^2\left[-\vec{\nabla_x}E+\eta(t)\right]}{m+\zeta\dt/2}
+O(\dt^4)
\end{eqnarray}
with $m$, $\zeta$, $E\equiv \rm H(\{\vec{x}\})$, $\eta$ as mass, viscosity,
energy (Hamiltonian), and uncorrelated thermal noise, respectively, where
$\{\vec{x}\}$ represents the set of coordinates for entire system.
Since the coordinates used in our simulations
represent C$_\alpha$ co-planes, the mass $m$ and the radius for viscosity
$\zeta$ are approximated from the mass of a single glutamine and
the C$_\alpha$ bond length, respectively.

The replica-exchange method \cite{replica} and multi-canonical rescaling
technique \cite{MC} are used to obtain the density of states $n(E)$ and
hence thermodynamic properties of any given system. First, several simulations
are performed at different temperatures. To enhance the sampling,
configurations obtained at different temperatures $T,T'$ were switched in
between based on the Metropolis criterion \cite{replica} that obeys detail
balance,
\begin{eqnarray}\label{replica-exchange}
P\left(\left[\{\vec{x}\}_{T},\{\vec{x'}\}_{T'}\right]
\rightarrow\left[\{\vec{x'}\}_{T},\{\vec{x}\}_{T'}\right]\right)
=\begin{cases}
1&\hspace{.5cm}\text{if $\Delta\le0$,}\\
e^{-\Delta}&\hspace{.5cm}\text{if $\Delta>0$.}\\
\end{cases}
\end{eqnarray}
with $\Delta=[1/T-1/T'][E(\{\vec{x'}\})-E(\{\vec{x}\})]$ (see
Ref. \onlinecite{replica} and references therein for more details).
Second, an initial
guess of $n(E)$ is obtained by using a WHAM (weighted
histogram analysis method)-like procedure \cite{Brooks,replica},
\begin{eqnarray}\label{WHAM}
P_\beta(E)&=&\frac{n(E)e^{-\beta E}}{\sum_E n(E)e^{-\beta E}}
\nonumber\\
\Rightarrow n(E)&\propto& \left[\sum_\beta
\frac{e^{-\beta E}}{P_\beta(E)}\right]^{-1}
\end{eqnarray}
where $\beta=1/k_BT$ ($k_B$ is Boltzmann factor) and $P_\beta(E)$ is
the probability of energy distribution accumulated from simulations at
temperature $T=1/k_B\beta$.

Then, we performed multi-canonical rescaling simulations to
refine $n(E)$, in which we used the same Hamiltonian but the
forces in equation of motion (\ref{Verlet-Langevin}) were rescaled as
$\vec{\nabla}E\rightarrow \frac{\partial E'(E,T)}{\partial E}\vec{\nabla}E$
with $E'(E,T)$ as a trial function. This rescaling will yield a probability
distribution $P_\beta(E)\propto n(E)e^{-\beta E'(E,T)}$; thus, if
we choose $E'(E,T)\sim k_BT\ln n(E)$, $P_\beta(E)$ will become relatively
flat and hence the sampling become more accurate (see \cite{MC} and
reference therein for detailed discussion).
Finally, the refined $n(E)$ is obtained iteratively,
\begin{eqnarray}\label{MC-iterate}
n(E)_{i}&\propto&
\left[\sum_\beta
\frac{e^{-\beta E'(E,1/\beta)_{i}}}{P_\beta(E)_{i}}
\right]^{-1}
=\left[\sum_\beta
\frac{e^{-\ln n(E)_{(i-1)}}}{P_\beta(E)_{i}}
\right]^{-1}
=\frac{n(E)_{(i-1)}}{\sum_\beta
\left[P_\beta(E)_{i}\right]^{-1}}
\end{eqnarray}
where $i$ indexes the iterations.

\subsection{The local melting approach}

Aside from examining the global cooperativity, we have also explored the
thermodynamics of local binding events in the model system.
The idea is that in the absence of hydrophobic clustering, the system
cooperativity (if any) must arise from a scaling up of small
scale bound structures.
In other words, folding of a specific part of the overall structure will be
aided by any contiguous folded regions \cite{Levine}; this can be
investigated by exploring the conditional probabilities of residues of being
folded (i.e., in their native position and orientation) or unfolded depending
on whether their neighbors are folded or unfolded (i.e., the contact
correlations).
Specifically, we investigated two cases; the ensemble that has all blocks
retained in their folded configurations aside from either a) a few blocks at
the end of one $\beta$-strand or b) a few blocks buried inside the sheet.

To study these local sub-structures,
we have assumed here that their thermodynamics
will not be affected significantly by the dynamics of very
distant blocks. This allows us to compute a ``conditional''
partition function in which, except for those specified blocks,
the rest can be treated as frozen in their native positions. Doing
this, the number of degrees of freedom is significantly reduced
and the conditional partition function can be obtained by explicit
numerical integration. For instance, in the single-block
$\beta$-tail case, the partition function reads $\int d{\rm X}_2
d{\rm X}_3 \delta(\overline{\rm X_1X_2}-\overline{\rm X_1X_2}^N)
\delta(\overline{\rm X_2X_3}-\overline{\rm X_2X_3}^N)e^{-H/k_BT}$
where $\overline{\rm X_1X_2}^N=3.75\stackrel{\circ}{\rm A}$,
$\overline{\rm X_2X_3}^N=0.57\stackrel{\circ}{\rm A}$ are native
constants, and ${\rm X}_2$, ${\rm X}_3$ are the virtual points of
the movable terminal block (note that its ${\rm X}_1$ is
fixed). Then, because of the $\delta$ functions, the partition
function is reduced to an integral over three angles defining the
orientation of the terminal plane.

\section{Results}

\subsection{Global Cooperativity}

The first system we studied is a 3-strand $\beta$-sheet with a
total of 6 H-bonds
equally distributed between adjacent strands. Since there is no sequence
specificity, we artificially tethered the H-bonds at the 2 expected turns
to maintain the $\beta$-sheet propensity. In Fig. \ref{nE1}, we show the
computed specific heat for the four different model forces. Basically, we
found that there is no sharp transition for the three models
${\rm LJ}^{fix}$, ${\rm LJ}^{ang}$, ${\rm Sol}^{fix}$ if non-specific H-bonds
are allowed to develop on the peptide backbone; this is because their
dominant low-energy states are ensembles with non-specific, randomly
collapsed structures rather than a uniquely defined $\beta$-sheet
(similar to the results observed by Guo \& Brooks \cite{Brooks}).
The complete model ${\rm Sol}^{ang}$, however, can define a
unique native state and give rise to a sharp transition, as seen in
Fig. \ref{nE1}(a).

Next, we added a \go-like restriction to the models
${\rm LJ}^{fix}$, ${\rm LJ}^{ang}$, ${\rm Sol}^{fix}$, i.e.,
allowing H-bond formation only between those residues that have
such an interaction in the native $\beta$-sheet structure \cite{Go}. Of
course, the entire justification of this approach is absent in
systems with homogeneous interaction as the one of primary concern
here, but it is still worthwhile comparing the results of our
proposed complete model with these alternatives. Note however that
our \go-like restriction does not include any additional
restrictions on backbone configurations to favor a particular structure
(i.e., native state); this is very different from the energetic restrictions
commonly used in the off-lattice protein folding literature \cite{cecilia}.
In Fig. \ref{nE1}(b), we show that even with \go-like restrictions
to reduce the non-specific collapsed ensemble, the
${\rm LJ}^{fix}$ model can not give rise to a sharp transition
compared with models incorporating  either the de/solvation effect
or angular dependency. This is consistent with results from other group
\cite{Berriz} and also the aforementioned reasoning that with only two
contact couplings for each interacting unit, a simple force field for
H-bond formation can not produce systemic cooperativity.

The difference between the various force models becomes much clearer in larger
systems. Fig. \ref{nE1}(c) shows the specific heat diagrams for a 4-strand
$\beta$-sheet with a total of 15 H-bonds equally distributed between adjacent
strands.
Now, without a significant increase in the transition temperature, the peak
magnitude of the specific heat for model ${\rm Sol}^{ang}$ is 100 fold
larger than that of the previous small system, whereas there is still no
significant transition for the L-J \go-like potential. This difference is
further
realized by examining the density of states. In Fig. \ref{nE2}(a),
we note that there is only one dominant ensemble for model ${\rm LJ}^{fix}$
in the low-energy region; this ensemble mixes the native and partially folded
states. In model ${\rm LJ}^{ang}$, on the other hand, the unfolded ensemble
($E\sim 0$) is slightly separated from the folded one; however, the folded
phase has a larger entropy than that of the unfolded state, and
thus there is no transition. In the solvation model, as shown in
Fig. \ref{nE2}(b),
there is a clear separation between the folded and unfolded ensemble,
with the unfolded states having the largest entropy, as is crucial for
having a sharp transition.

Apparently, both the angular dependence and solvation effect slightly enhance
the system cooperativity, but only a combination of both can yield the
desired results in the system of primary concern here. This point has not been
adequately addressed in previous works \cite{Berriz,Brooks}. Also, we notice a
high energy population appearing in model ${\rm LJ}^{ang}$, ${\rm Sol}^{fix}$,
and ${\rm Sol}^{ang}$. Analyzing the contact profile in this additional
phase, we found that its dominant configurations are those with significant
intervening of folding and unfolding in the entire structure, i.e.,
``droplet''-like structures \cite{Levine,wolynes}. As will be
elucidated in next section, this intervening can lead to an interfacial
energetic penalty resulting from backbone twisting (enhanced by the angular
dependence) and (re)-desolvation barrier crossing at the interface between
contiguous folded and unfolded regions; thus the partially unfolded ensemble is
shifted from low energy (as in model ${\rm LJ}^{fix}$) to the high energy
region. Consequently, the completely folded and unfolded ensemble are well
separated, leading to a sharp transition.

Of course, whether these partially folded droplet configurations
can be shifted to high energy depends not only on the choice of
the H-bond potential (to create the interfacial penalty), but
also the topology of the $\beta$-sheet. When the H-bond energy
inside the folded droplet can compensate the interfacial penalty,
the structure will no longer stay at high-energy phase. To show
this explicitly, we studied a system that is less symmetric on the
$\beta$-sheet plane; we chose a 3-strand $\beta$-sheet with long
strands, looking at the effect of increasing the strand length. We
started with a 3-strand $\beta$-sheet with a total of 30 H-bonds
equally distributed between adjacent strands. Fig. \ref{nE3}(a,b)
shows its $n(E)$ and specific heat, which are very similar to
those of the 4-strand, 15 H-bond system, in that they both have
well-separated unfolded/native states, and partially unfolded
high-energy populations. Analyzing the states in the low and high
energy partially folded ensembles, we found that their dominant
configurations are droplets with one (in the low energy phase) or
multiple (in the high energy phase) interfaces between contiguous
un/folded regions, i.e., a partially folded $\beta$-sheet (or
hairpin) buried in unfolded coils (illustrated by examples in Fig.
\ref{nE3}(a)).

When the system size increases to a total of 40 H-bonds, however,
(see Fig.\ref{nE4}(a,b)), we find an increased weight for the
partially folded ensemble in the low energy phase ranging from $0
>E > E_{N}$ where $E_N$ is the native state energy.  Unlike the smaller, or
symmetric system, here the interfacial energy penalty between
contiguous un/folded regions, can not compete with the large
H-bond energy at the bulky folded portion; this allows the
partially folded ensemble to have a low energy. Moreover, in a
system with such an elongated asymmetry, the partially folded
portion can freely move along the $\beta$-sheet, or freely
slide along the chain with many ``mis-pairing'' H-bonds (i.e.,
inconsistent with native pairing). Both effects can then increase
the entropy of the partially folded (or even misfolded) ensemble
and hence smears out the separation between completely
un/folded ensemble and partially folded ensemble, Fig.
\ref{nE4}(a). The smearing becomes more significant in a
larger 3-strand $\beta$-sheet with a total of 50 H-bonds, Fig.
\ref{nE4}(c). As a consequence, the system becomes less
cooperative and an additional continuous transition occurs
between these partially un/folded ensemble, fig.\ref{nE4}(b,c).
Developing a phenomenological model based on these results to
study how the system cooperativity depends on the $\beta$-sheet
topology, therefore, is one of our future goals.

\subsection{Local Bonding Effects}

In this section, we examine how the model system acquires ``global''
cooperativity (in the absence of hydrophobic clustering) by exploring
the thermodynamic effects of local un/bonding.
Here, since we are dealing with only a small number of degrees
of freedom, we could calculate all thermodynamic properties by explicit
integration of the partition function. As an illustration, we first present
the results of the single-block $\beta$-tail case, Fig. \ref{1-block}, where
we plotted the probability distribution function (pdf) $P(r)$ of the
interacting distance $r$ in the native NH$\cdots$OC pair between the terminal
block (the one allowed to fluctuate) and its native partner
(taken to be frozen in space).

In this case, the maximum possible NH$\cdots$OC distance as determined
by the range of integration, is smaller than that necessary to fit a water
molecule in between and thus there cannot be any true secondary solvated
minimum. We do see however the fact that having the solvation barrier and
(to a lesser extent) the angular factors make a big difference in the
structure of $P(r)$. Specifically, there is no hint of well-separated 2-state
behavior for model ${\rm LJ}^{fix}$. However, a small barrier appears if the
potential has an angular dependence (model ${\rm LJ}^{ang}$), and becomes
significant if solvation effects are included (model ${\rm Sol}^{fix}$,
${\rm Sol}^{ang}$). Here, we note that because of the bond angle
$\rm \angle C'C_\alpha N$ potential, $P(r)$ is concentrated on the ``ring''
that satisfies $\rm \angle C' C_\alpha N=\angle C' C_\alpha N_{native}$. On
this ring, only a few ``points'' will give minimal dihedral angle
($\psi$, $\phi$) energies. These points correspond to the sharp peaks in
fig.\ref{1-block} that have free energy difference $\approx
\epsilon_{\psi}+\epsilon_{\phi} \approx 1 \rm kcal/mol$ from their
surrounding; if we reduce the bond angle stiffness $k_{\theta}$,
these sharp peaks become broadened (data not shown). Apparently,
this effect comes from the backbone energy and is independent
of the choice of H-bond potential.

Next, we studied the``buried'' cases with $n$ blocks allowed to
fluctuate. In this situation, the configurations of fluctuating
blocks are clearly constrained to a great extent by their fixed
neighbors. Thus, we found that for $n=1$, there is neither 2-state
behavior nor any qualitative difference among the results of
different force fields (data not shown); all significant
probability is confined in the H-bond well, indicating a strong
topological confinement to force the folding of a single unfolded
block buried amidst folded ones. When $n=2$, we still do not have
enough freedom of backbone motion to allow for a solvated second
minimum for the main H-bond; the maximum separation is slightly
above 4.1 $\stackrel{\circ}{\rm A}$, Fig. \ref{2-block}(a).
Nonetheless, we see clear differences among different models. For
the simplest LJ$^{fix}$ model, there is no confinement to a narrow
contact well. Adding the angular factor back in, we create a
barrier as the NH$\cdots$OC distance is increased. This barrier is due
to the modulation of the LJ potential by the angular factor,
making it repulsive if the orientation is not properly matched.
Alternatively, adding the solvation effect as in model Sol$^{fix}$
also introduces a barrier; this is because rolling of the co-plane
allows for a wide separation (and hence going up and down the
solvation barrier) in the C$_\alpha$H$\cdots$OC interactions. Finally,
putting both effects in leads to a very large degree of
confinement in the H-bond well, suggesting that local coupling is
enhanced by both structural and solvation effects such that an
unfolded block is strongly biased to fold if contiguous to folded
ones.

To further characterize the difference between the varying choices
of force models, we explored the un/folding correlation in the 2-block
buried case. Specifically, we computed the conditional probability function
$P(r)$ for one block with respect to the un/folding state of another (defined
as folded/unfolded if its native H-bond separation $r_2\gtrless\rb$),
$P(r_1|r_2>\rb),P(r_1|r_2<\rb)$, where $r_1$, $r_2$ are their native H-bond
separations, respectively. In other words, if the system has a good local
coupling, un/folding of one block should force the un/folding of another,
implying a well-separated $P(r_1|r_2\gtrless\rb)$. In Fig. \ref{2-block}(b),
however, we found that $P(r_1|r_2\gtrless\rb)$ are well separated
only in model Sol$^{ang}$. Here, the separated probability distributions
 have been
individually normalized; the actual probability for being at the larger $r$
is extremely small as is evident from Fig. \ref{2-block}(a).
Also, note that it is not $r$ in the figure which is important - this
H-bond separation is quite small. Instead, the two peaks
correspond to either having the full H-bond energy (by satisfying {\em both}
the distance and angular conditions) or not having that energy; the
distribution at larger $r$ does not have the correct angular configuration.
To clarify this point, we plotted the conditional probability function
on the variation of H-bond energy $E$ instead of distance $r$,
i.e., $P(E_1(r_1)|r_2>\rb),P(E_1(r_1)|r_2<\rb)$, Fig. \ref{2-block}(b).
Now, we see clearly that only the full model Sol$^{ang}$ has a significant
separation effect.

\subsection{Collective Re-solvation Kinetics}

The conclusion from the local coupling studies is that the full
model being proposed here leads to strong correlations between the
behaviors of neighboring blocks. For the system as a whole, this
should mean that complicated configurations with repeated
interfaces between unfolded and folded residues would cause
additional energetic penalties and thus can be strongly suppressed in
favor of simpler configurations with large patches of residues being
either all in contact or all not in contact. Of course, whether the
system can therefore acquire an global cooperativity
would depend on its topology as well, as revealed in earlier section.

Now, the interesting part of this conclusion, as will be illustrated
later, is that during unfolding (folding), the solvation (desolvation)
process must occur in a collective way.
To see this explicitly, we show the pdf and re-solvation kinetics of a
10-block buried case in  Fig. \ref{kinetics}(a,b). Because of the large barrier
needed to melt the blocks, here we set the simulation temperature far beyond
the physiological range.

From Fig. \ref{kinetics}(a) (using Sol$^{ang}$), we note that the pdf
of native H-bond distances are well separated by a free energy (-$k_BT\ln
P(r)$) barrier, resulting in an ordered (H-bond formed) and
disordered (H-bond not formed) phases. Interestingly, even
governed by the same force law, the un/folding barriers for the
various molten blocks are not quite the same. For blocks close to
the confined ones (those frozen in the space), the barrier
progressively increases, and the disordered phase becomes less
significant. This reduction of disordered phase manifests a
stronger folding bias for any unfolded residues near those folded
ones.  Thus it is not effective to unbind the blocks one by one,
since the unbound block will rebind long before its neighbor has
unbound.  This can be seen in the high-temperature unfolding
simulation, Fig. \ref{kinetics}(b), in which we note that the
resolvation (i.e., infiltration of water molecules) process
involved many residues simultaneously. In Fig. \ref{kinetics}(c),
we show that this collective water infiltration indeed occurs in
unfolding of the entire $\beta$-sheet.

\section{Discussion} 

We have explored the mechanism of $\beta$-sheet cooperativity
without side-chain interactions, with different choices of the
H-bond potential. Our results suggest that in the absence of
hydrophobic clustering, the $\beta$-sheet cooperativity arises
from a coupling of structural factor (contact angular dependence),
protein de/solvation effects, and the system topology. Since this
phenomenon is sequence-independent, the results can be considered
to apply in general to $\beta$-sheet structures.

Given the coupling number for each H-bond interacting unit, our
results differ from most 2-dimensional pairwise-interacting system,
in that there the minimal number of coupling required for global cooperativity
is larger than the one present here. Thus, the coupling between
backbone energy (the angular dependence) and pairwise H-bond interaction
(incorporated with solvation dynamics) must yield a strong many-body effect,
which has been argued to be crucial to produce systemic cooperativity
in homogeneous system such as are the primary concern here \cite{Levine}.
Apparently, our model system has a default many-body effect,
the dihedral confinement via the corresponding terms in the backbone energy.

As illustrated in our study, when the system is close to the
fully folded state, the dihedral confinement is strong enough to complete the
folding (the $n=1$-block case). However, it is too weak to confine the
backbone collectively if there is a larger unfolded portion in the
$\beta$-sheet, as illustrated in the pure Lennard-Jones potential case.
Instead, it is the angular dependence which incorporates the
contact strength into the backbone Hamiltonian and thus amplifies the
confinement. This leads to a competition with solvation dynamics; in other
words, while the solvent tries to pull the interacting blocks
apart, the confining force amplified by folded neighbors can hold
them back. As this behavior occurs mainly between contiguous un/folded
regions, it creates an ``interfacial'' energetic penalty and the ``droplet''
picture proposed in previous works \cite{Levine,wolynes} (now allowing
a measurement the droplet surface tension).
Consequently, in a partially folded structure, any
small regions of unfolded blocks contiguous to folded ones are
thermodynamically unfavorable and will be forced to fold or will
unfold the folded ones \cite{Levine}.
Thus, while there has been a concentration of effort on understanding
the role the sequence heterogeneity for protein folding and
design, we would like to point out that this generic cooperativity
buried in the $\beta$-sheet architecture can also serve as a
{\em building block} to construct macromolecular structure. However,
this same building block can also provides the mechanism for
pathological aggregation.

What then would be the physiological role and evolutionary
principle of having sequence heterogeneity, in the presence of
possible $\beta$-architectures? Certainly, with the fast time
scale (of order of $0.1$ ns, estimated from our Langevin
simulations) for a single C$_\alpha$ block folding, any pre-folded
$\beta$-architecture with ordered, exposed amide and carbonyl
oxygen could lead to a seeding process and non-specifically
precipitate surrounding unfolded peptides; this has been observed
in the elongation process of amyloidogenesis \cite{Kelly2}. The
physiological role of heterogeneity, therefore, would be to either
distort the orientation of exposed amide and carbonyl groups into
a non-aggregation competent form, or provide ``hot'' nuclei to bias
the system into the native fold. Indeed, this idea has been
verified in a recent tryptophan zipper model \cite{cochran}.
Exploring the competition between native folding due to specific
sequence heterogeneity and pathological aggregation due to generic
cooperativity is perhaps the most important direction for future
study.

As a side point, to elucidate the effect of local un/binding, we have
introduced a local melting method. This method can be further used to
explore local nucleation or de-solvation events in any system
lacking long-ranged spatial couplings.
Also, instead of simulating all backbone atoms or a
simplified C$_\alpha$ model, a functional block (co-plane)
approach was proposed. This approach greatly reduces our
computational effort; however, it still keeps the right angular
degrees of freedom (missing in C$_\alpha$ model) and gives rise
to a structural factor (i.e., backbone alignment) that helps account
for $\beta$-sheet cooperativity.

Also, we have introduced an approach to de/solvation
which can be used directly for Langevin simulations. It has been
noticed that the (de)-solvation effects are important in protein
folding and functioning, but no standard model has been proposed
\cite{Hummer,water,Dunitz,tension,Helson,Stillinger,VWF1,solvent}.
Also, we note that solvent dynamics can be regulated by  system
hydrostatic pressure \cite{Hummer}. Interestingly, many adhesion and
hemostasis-related proteins are very sensitive to shear stress and
pressure change; perhaps a local de/re-solvation event might
trigger a cooperative conformational change which ultimately has
large scale consequences \cite{VWF1}. Thus, microscopic
cooperativity can be amplified into a macroscopic response; this
also needs to be explored in future work.

\section{Acknowledgement}

CG wishes to acknowledge Peter G. Wolynes, Jose N. Onuchic, Angel
E. Garcia, Tobin R. Sosnick, Joan-Emma Shea, and Ken A. Dill for
comments the solvation model, and Christopher M. Dobson for
comments on sequence-independent mechanism. He also thanks to A.E.
Garcia for a careful reading on the manuscript.

\newpage

\section{Figure Captions}

{\bf Fig.1}
{\bf(a)} The co-planar structure with {\bf(b)} the
three virtual points $\rm\{X_1,X_2,X_3\}$ and dihedral angles
$\omega,\phi,\psi$ labeled. The virtual bond
lengths and angles are determined so as to give {\bf(c)} a
regular, symmetric $\beta$-sheet structure with side-chains and
C$_\alpha$H bonds alternately pointing up and down
periodically. {\bf(d)} The side-chain and C$_\alpha$-H$\cdots$OC
interactions are indicated by light-gray, gray arrows, respectively,
with the shadowed area indicating the atoms involved determining
the angular dependence of a single H-bond. {\bf(e)} If two
interacting blocks $i,j$ are to have a native H-bond (the gray
arrow), proper alignment of two pairs of planar vectors
and collinearity of the two sets of points
(X$_1^{(i)}$,X$_2^{(j)}$,X$_3^{(j)}$;
X$_1^{(j)}$,X$_3^{(i)}$,X$_2^{(i)}$) are required.

{\bf Fig.2}
The structural-factor-dependent H-bond potential profiles in units of
kcal/mol with interaction distance $r$ (NH$\cdots$OC) measured in
$\stackrel{\circ}{\rm A}$. {\bf(a)} The (de)-solvation force field
${\rm U_{Hb}}(r,\x)$. The barrier height between $r=\rb$ and the
solvent-separated minimum ($r=\rc$) represents the free energy
cost in desolvation cavity formation. Here we have graphed ${\rm
U_{Hb}}(r,\x)$ from $\x=0$ to 1. The dashed curve represents
the modulation of the contact minimum $(\rx,{\rm
U_{Hb}}(\rx,\x))$. {\bf(b)} The Lennard-Jones force field ${\rm
U}_{\rm LJ}^{ang}(r)$, where the profile is modulated by changing
the contact minimum (long-dashed curve) $|{\rm U}_{\rm LJ}^{ang}(r_{\rm
LJ}(\x))| =\x|E_{\rm Hb}|+(1-\x)|E^{sol}|$ from the H-bond contact
energy $E_{\rm Hb}$ to the single H$_2$O hydration $E^{sol}$. The
major difference between {\bf(a)} and {\bf(b)} is the desolvation
barrier.

{\bf Fig.3}
{\bf(a)} The specific heat for a 3-strand with 6 H-bonds modeled
by different force fields (without \go-like restrictions).
Here, because of the absence of \go-like pairing, the three
force field ${\rm LJ}^{fix}$, ${\rm LJ}^{ang}$, and ${\rm Sol}^{fix}$
are unable to define a unique structure and hence there is no obvious
peak (i.e., transition) in their CV diagrams. Whereas force field
${\rm Sol}^{ang}$ can define unique $\beta$-sheet and hence leads to
a sharper transition.
{\bf(b)} Same as {\bf(a)} except for the \go-like restrictions
to model ${\rm LJ}^{fix}$, ${\rm LJ}^{ang}$, and ${\rm Sol}^{fix}$ here.
Note that the transitions for these three force fields
become more apparent but still not compatible with ${\rm Sol}^{ang}$.
{\bf(c)} The specific heat for a 4-strand with 15 H-bonds modeled
by different force fields.
Now the transition for ${\rm Sol}^{ang}$ become much stronger and there is
almost no transition for L-J type force fields (inset).

{\bf Fig.4}
The density of state for a 4-strand $\beta$-sheet with 15 H-bonds.
{\bf (a):} The density of state for L-J type potential (with \go-like
restrictions). Note the distribution of entropy and the high-energy population
appearing in model ${\rm LJ}^{ang}$; this population contains
partially folded state.
{\bf (b):} The density of state for Sol type potential. Note that a good
separation between the unfolded and folded ensemble only appeared in model
${\rm Sol}^{ang}$.

{\bf Fig.5}
The density of state {\bf (a):} $n(E)$, specific heat {\bf (b)}
$C_v=\frac{\partial\langle E\rangle}{\partial T}$, and
energy $\langle E\rangle$ (inset in {\bf (b)})
for system asymmetrically elongated at one direction on its $\beta$-sheet
plane (using model Sol$^{ang}$). Here the system is a 3-strand $\beta$-sheet
with total 30 H-bonds equally distributed between adjacent strands.
Note the suppression of low-energy ensemble (between the completely unfolded
and native states) and the appearance of high-energy population.
The dominant configurations in these ensemble (insets) are ``droplets'' with
interfaces between contiguous un/folded regions. Note that the low-energy
configuration has less interface compared with high-energy one. The negative
energy of the unfolded state comes from the two tethered $\beta$-turns.

{\bf Fig.6}
{\bf (a):} The $n(E)$ for a 3-strand $\beta$-sheet with total 40
H-bonds. Note the growth of partially folded
ensemble (inset in {\bf (a)} is a typical example of their configurations)
at low-energy phase and smears out the separation between unfolded
and the native states. This can lead to an abrupt transition (with the
native state) at low temperature phase (pointed by double arrows in {\bf (b)})
and a continuous transition (with other high-energy unfolded states) at
high temperature phase (pointed by a single arrow in {\bf (b)}).
When the system size increases to totally 50 H-bonds {\bf (c)}, the
smearing becomes more significant (arrow in {\bf (c)}) and the system
eventually loses its cooperativity.

{\bf Fig.7}
The numerical results for the probability distribution function (pdf) $P(r)$
of NH$\cdots$OC separation $r$ of the ``molten'' block respective to its
native partner in the $\beta$-tail single-block case (indicated in the
inserted diagram; black: frozen, gray: mobile).
Here temperature: 25$^{\circ}$C and integration step $\delta\theta=10^{-4}$.
Also, the curves have been adjusted up/downward for a better visual comparison.
Note that the barrier for model Sol$^{fix}$ is much larger than that of
Sol$^{ang}$. This is because the desolvation barrier height is fixed in the
former case, but modulated (by the angular factor) in the latter one.

{\bf Fig.8}
The numerical results for buried two-block cases.
{\bf(a)} The pdf $P(r)$ for the native H-bond separation (average of the two
symmetric mobile blocks).
{\bf(b)} The normalized conditional pdf in first molten block, $P(r_1)$, with
respect to whether the second block is unfolded (defined as
if $r_2\ge \rb$) or folded (defined as if $r_2< \rb$). Here $r_1$,
$r_2$ are the H-bond separations of the two molten blocks with their
native partner, respectively, and $\int dr_1 P(r_1|r_2\gtrless\rb)=1$.
{\bf(c)} The normalized conditional pdf in first molten block, $P(E_1)$,
plotted on its H-bond energy $E_1$.

{\bf Fig.9} {\bf (a):} Langevin simulation results of a ten-block
buried case. Each curve represents the corresponding block in the
inset diagram, averaged over the symmetrical pair. The simulation
is averaged over 25 series of 1 ms runs ($\dt=5\times 10^{-5}$
ps). {\bf (b):} Snapshot of re-solvation simulation at $\rm
T=850^{\circ}K$ where each heavy atom is labeled by different
symbols depending on the corresponding H-bond interacting distance
$r$: ``$\bullet$'' ($r<\rb$), ``$\circ$'' ($\rb\le r<\rc$), and ``
'' ($r\ge\rc$). {\bf (c):} Snapshot of unfolding simulation of a
4-strand $\beta$-sheet with 15 H-bonds at $\rm T=800^{\circ}K$
where we have tethered the H-bond at the turn area to maintain the
$\beta$-turn propensity. The time courses are plotted on the
change of total H-bond (indicated by $r\le\rb$) number $\rm
Q_{Hb}$ and energy $\rm E_{Hb}$ (kcal/mol), and configurations at
different snapshots (a,b,c,d,e) are projected to left-hand side.
Note the successive infiltration of water, and the overshot of
$\rm E_{Hb}$ at the abrupt transition point, manifesting a
collective barrier crossing for the re-solvation process.

\appendix
\section{Modified desolvation potential}\label{appen}

\begin{figure}
\centerline{\hbox{ {\epsfxsize = 3.4in \epsffile{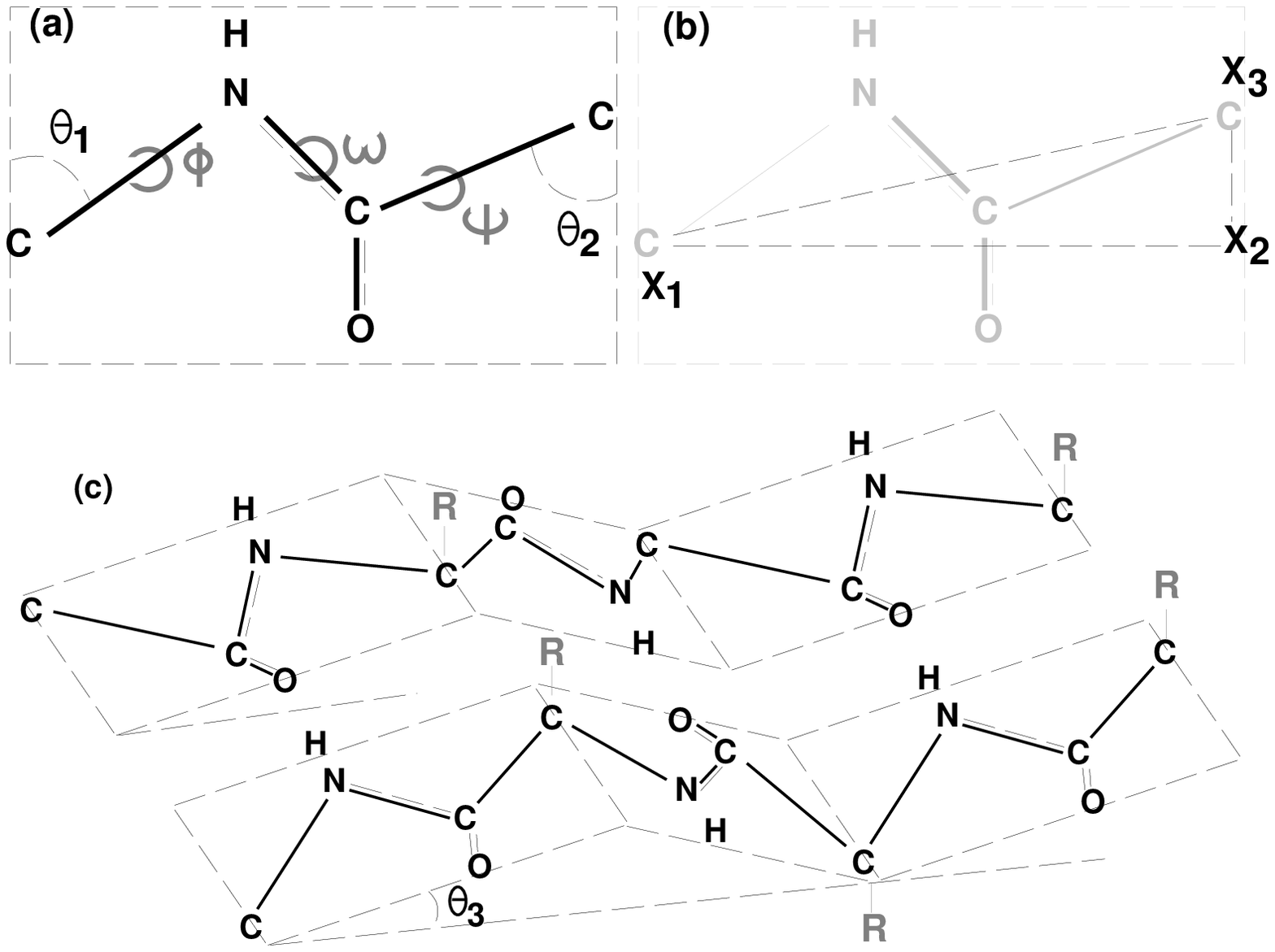}}
{\epsfxsize = 3.0in \epsffile{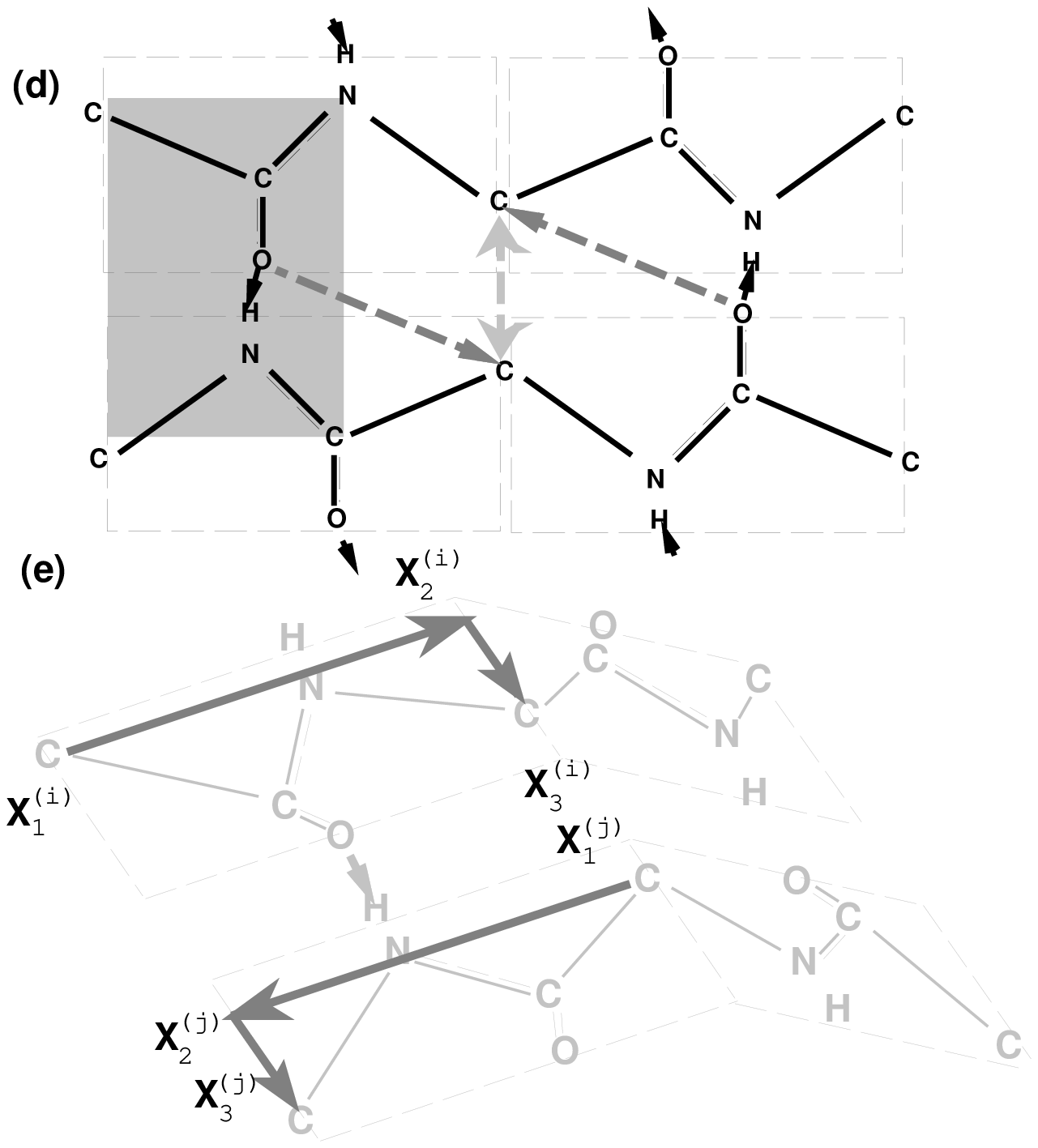}} }}
\caption{}\label{fig0}
\end{figure}

\begin{figure}
\centerline{\hbox{
{\epsfxsize = 3.2in \epsffile{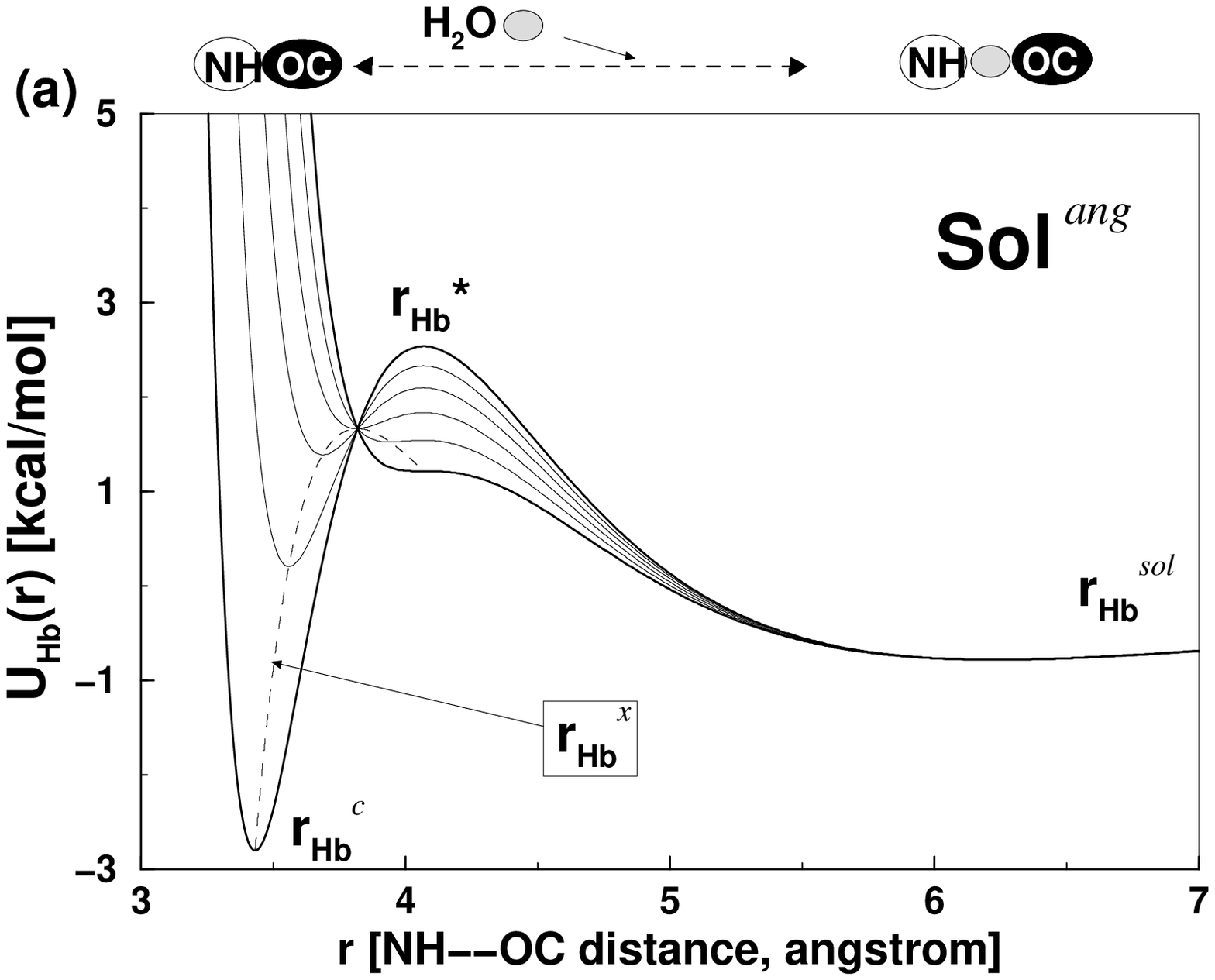}}
{\epsfxsize = 3.2in \epsffile{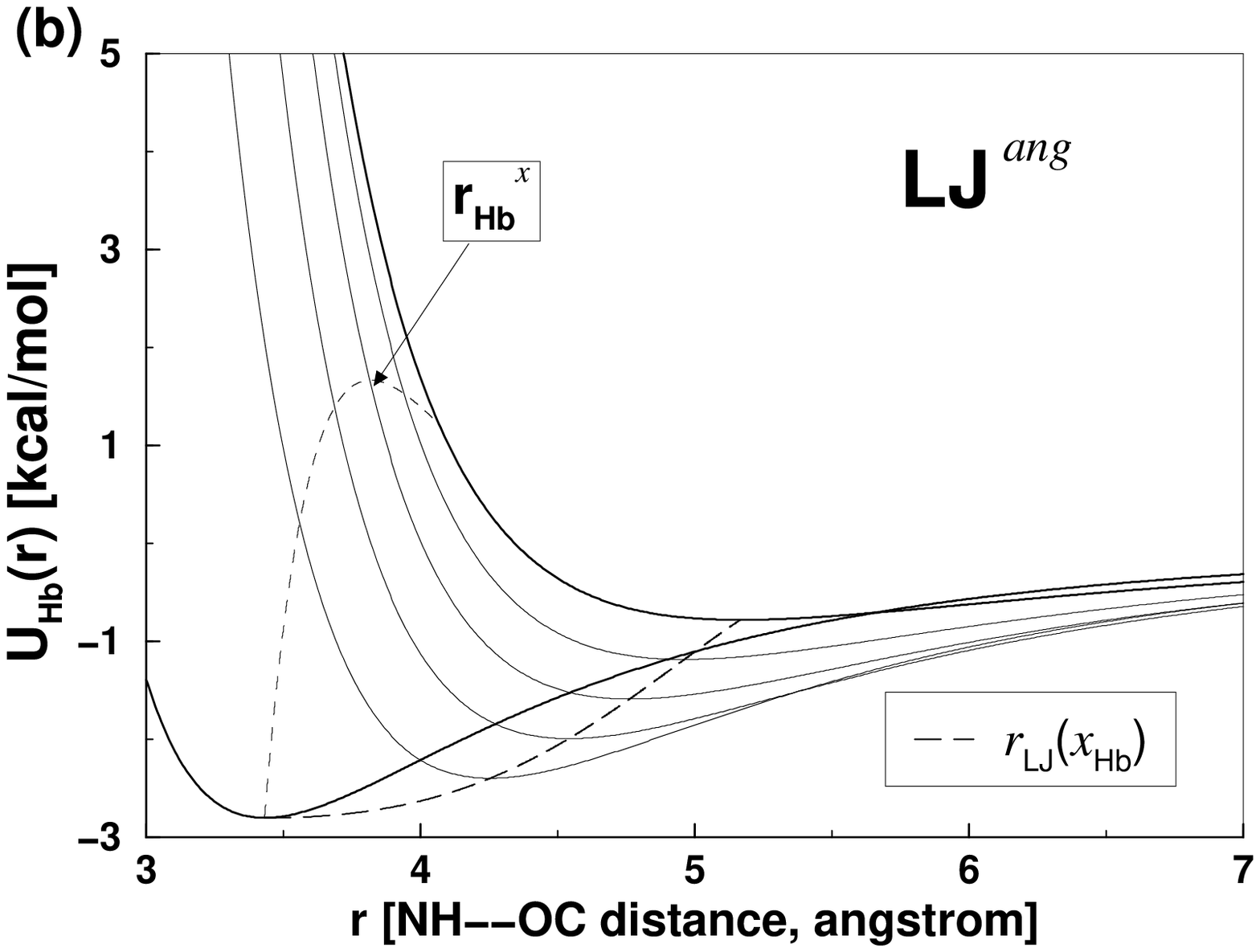}}
}}
\vspace{.5cm} \caption{}\label{fig1}
\end{figure}

\begin{figure}
\centerline{\hbox{
{\epsfxsize = 3.2in \epsffile{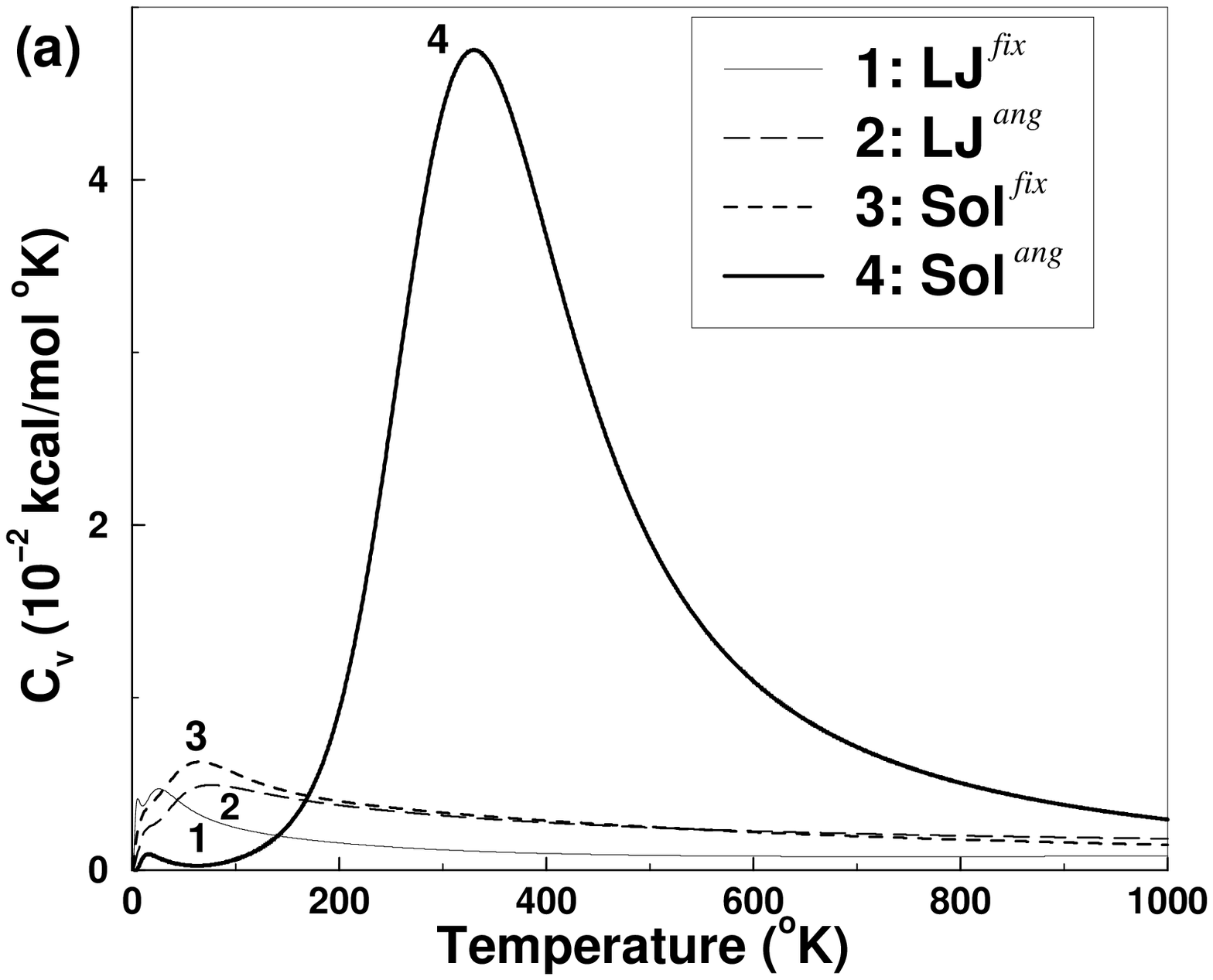}}
{\epsfxsize = 3.2in \epsffile{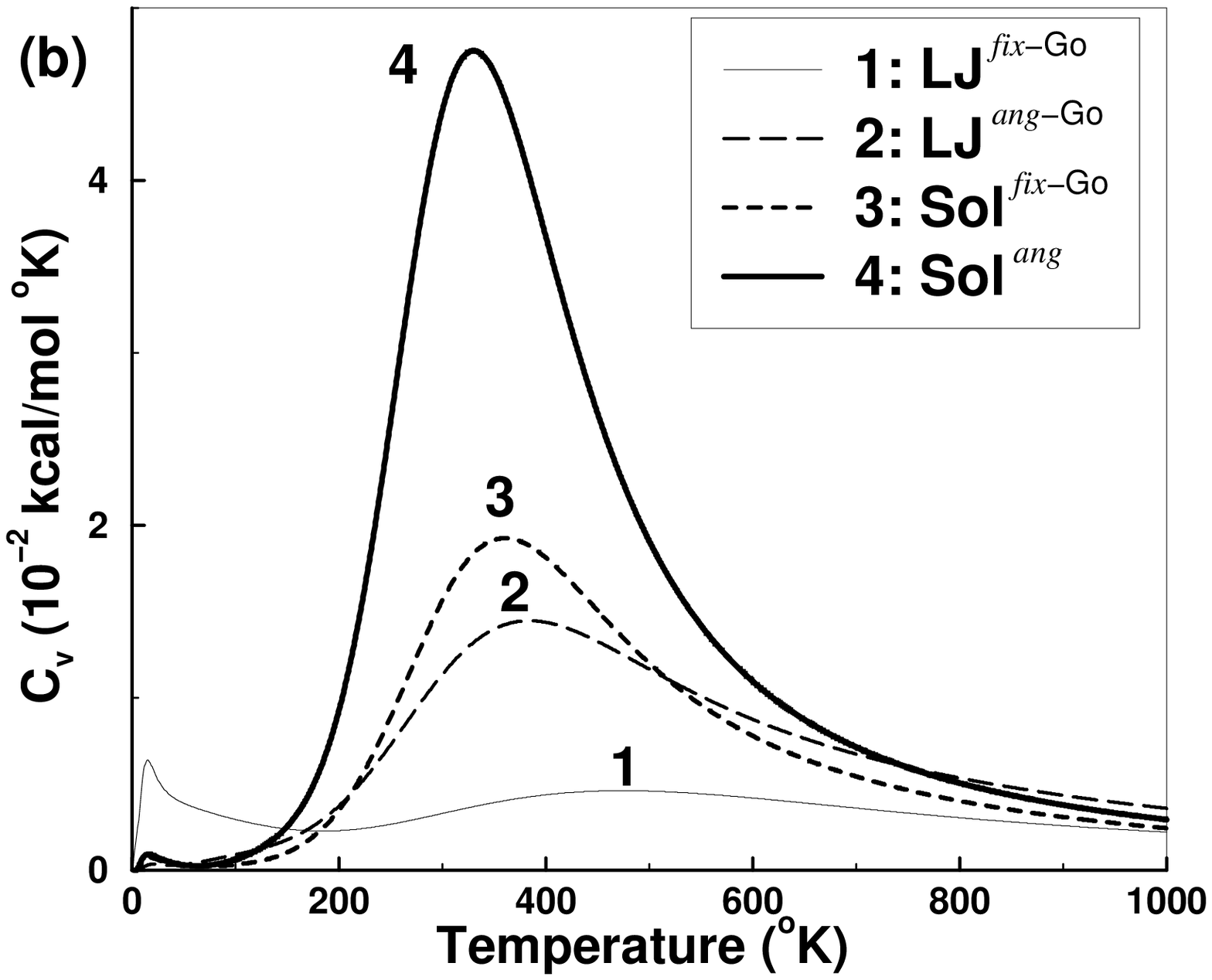}}
}}
\vspace{.5cm}
\centerline{
{\epsfxsize = 3.2in \epsffile{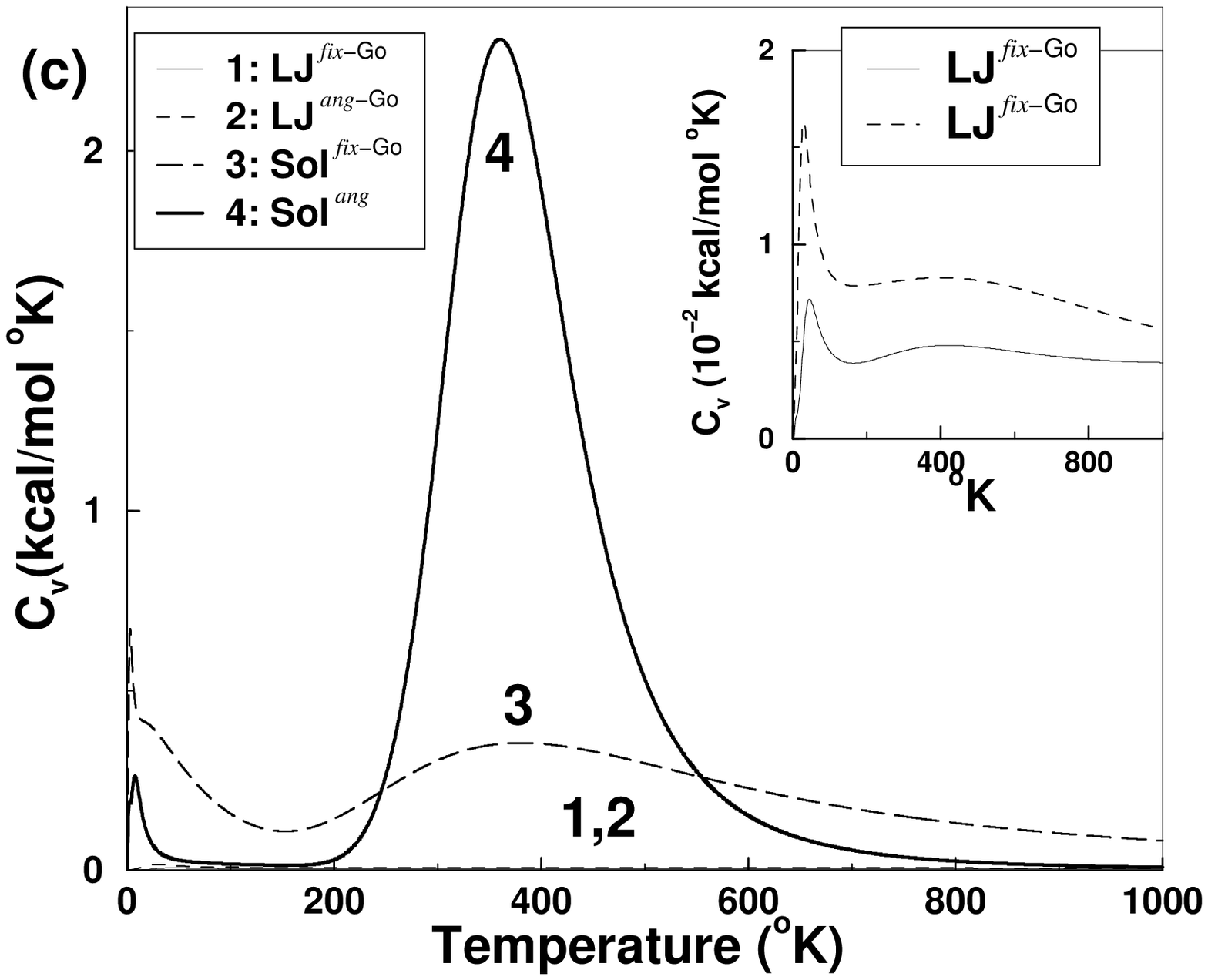}}
}
\vspace{.5cm} \caption{}\label{nE1}
\end{figure}

\begin{figure}
\centerline{\hbox{
{\epsfxsize = 3.2in \epsffile{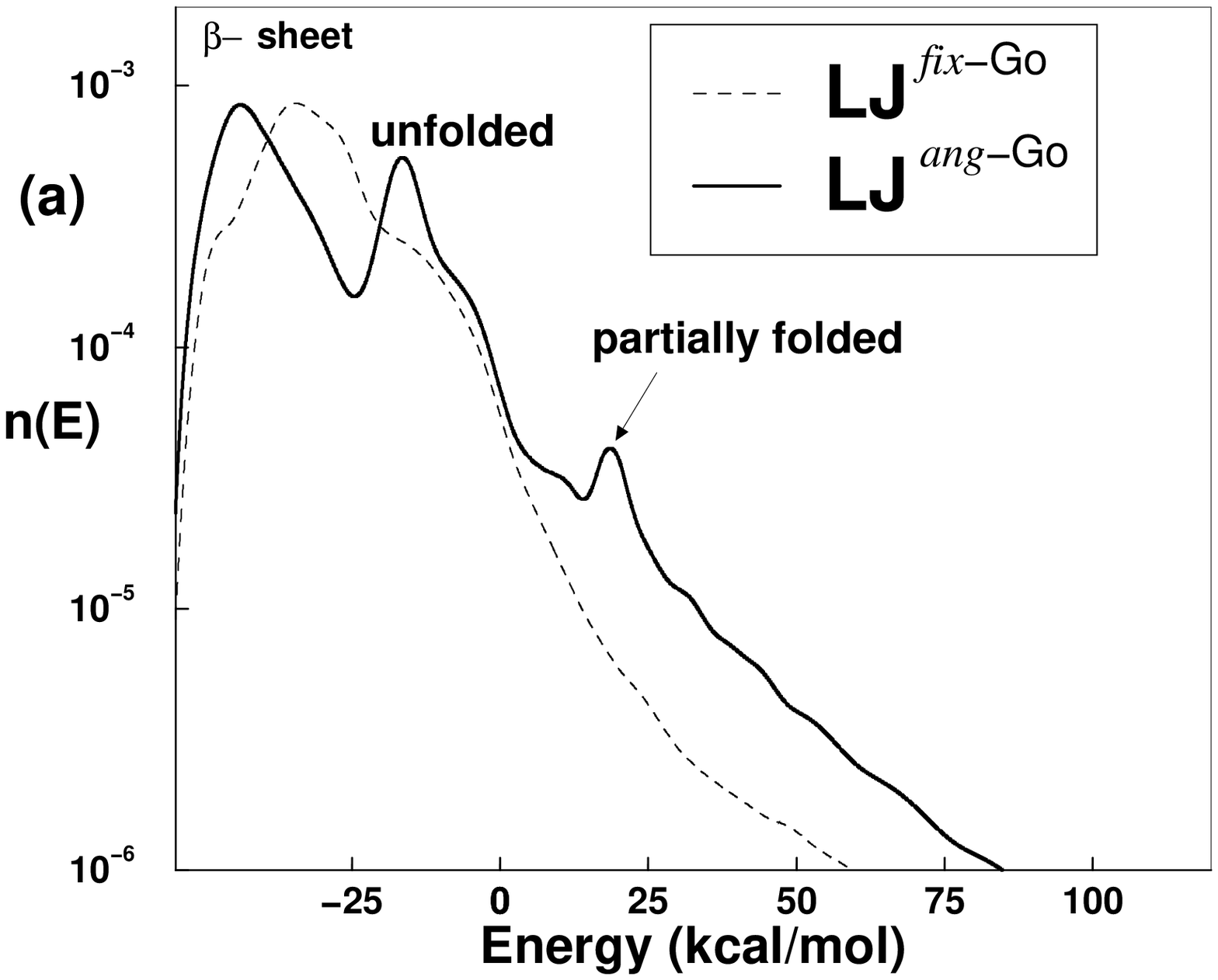}}
{\epsfxsize = 3.2in \epsffile{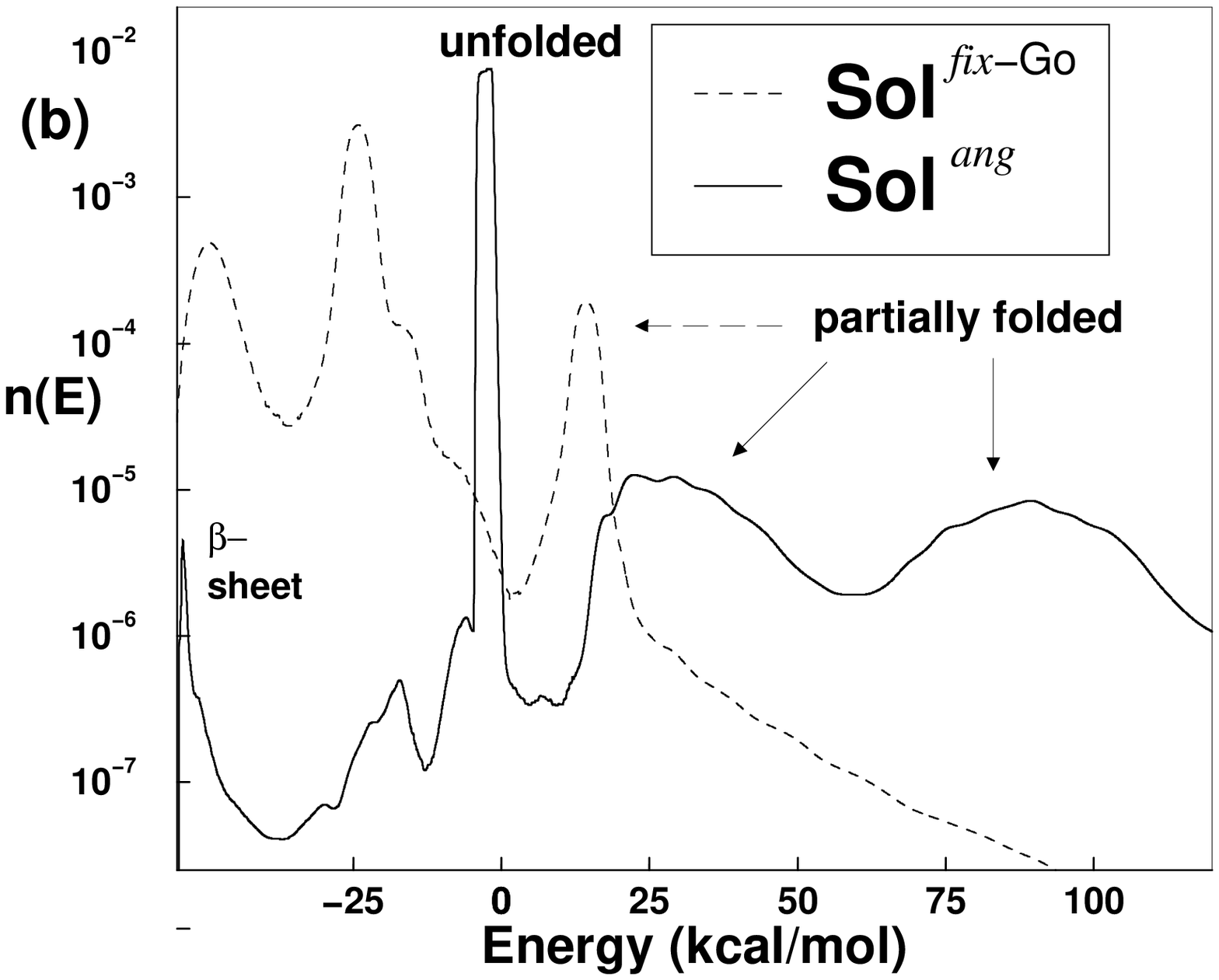}}
}}
\vspace{0.5cm}
\caption{}\label{nE2}
\end{figure}

\begin{figure}
\centerline{\hbox{
{\epsfxsize = 3.15in \epsffile{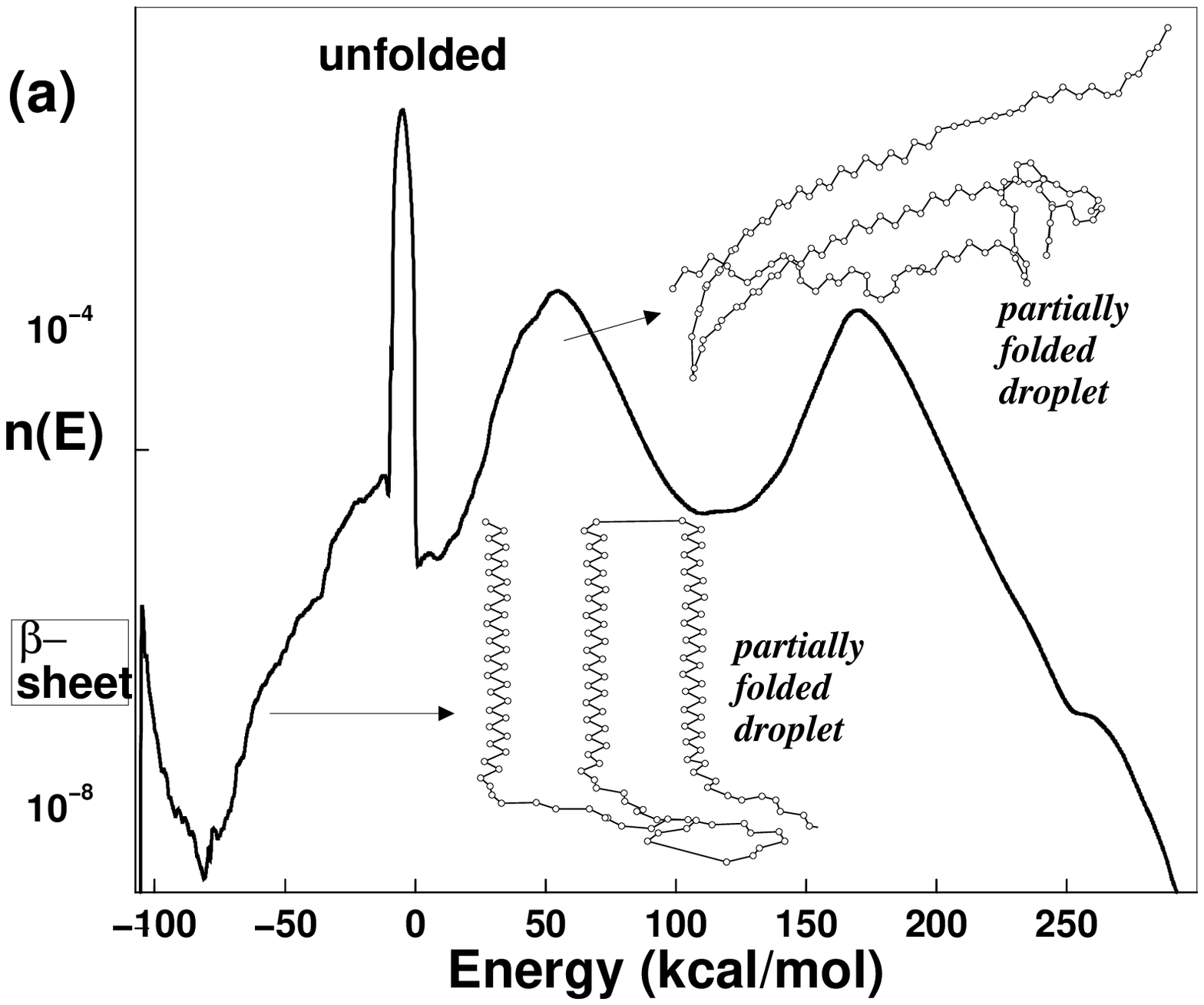}}
{\epsfxsize = 3.2in \epsffile{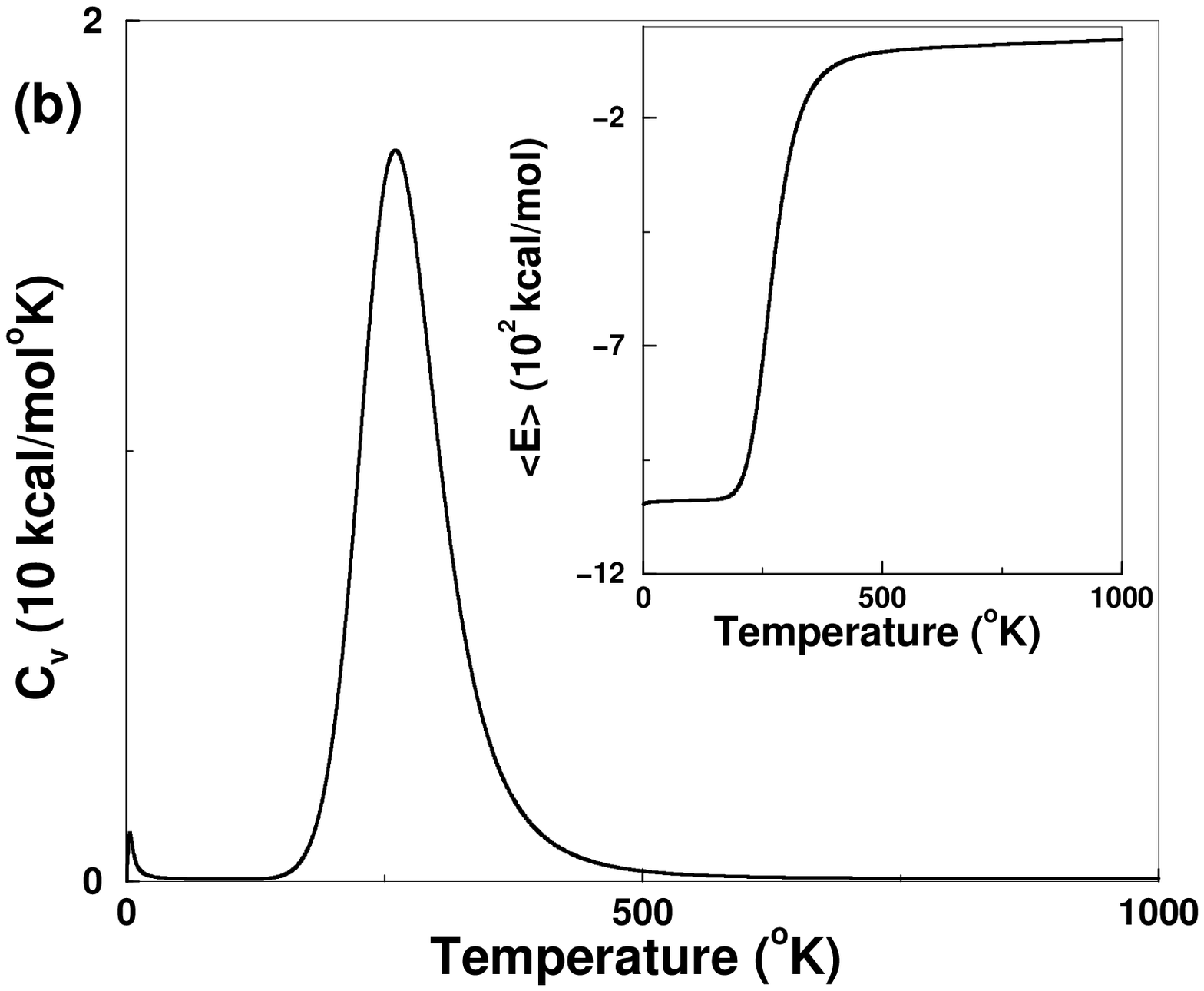}}
}}
\vspace{0.5cm}
\caption{}\label{nE3}
\end{figure}

\begin{figure}
\centerline{\hbox{
{\epsfxsize = 3.15in \epsffile{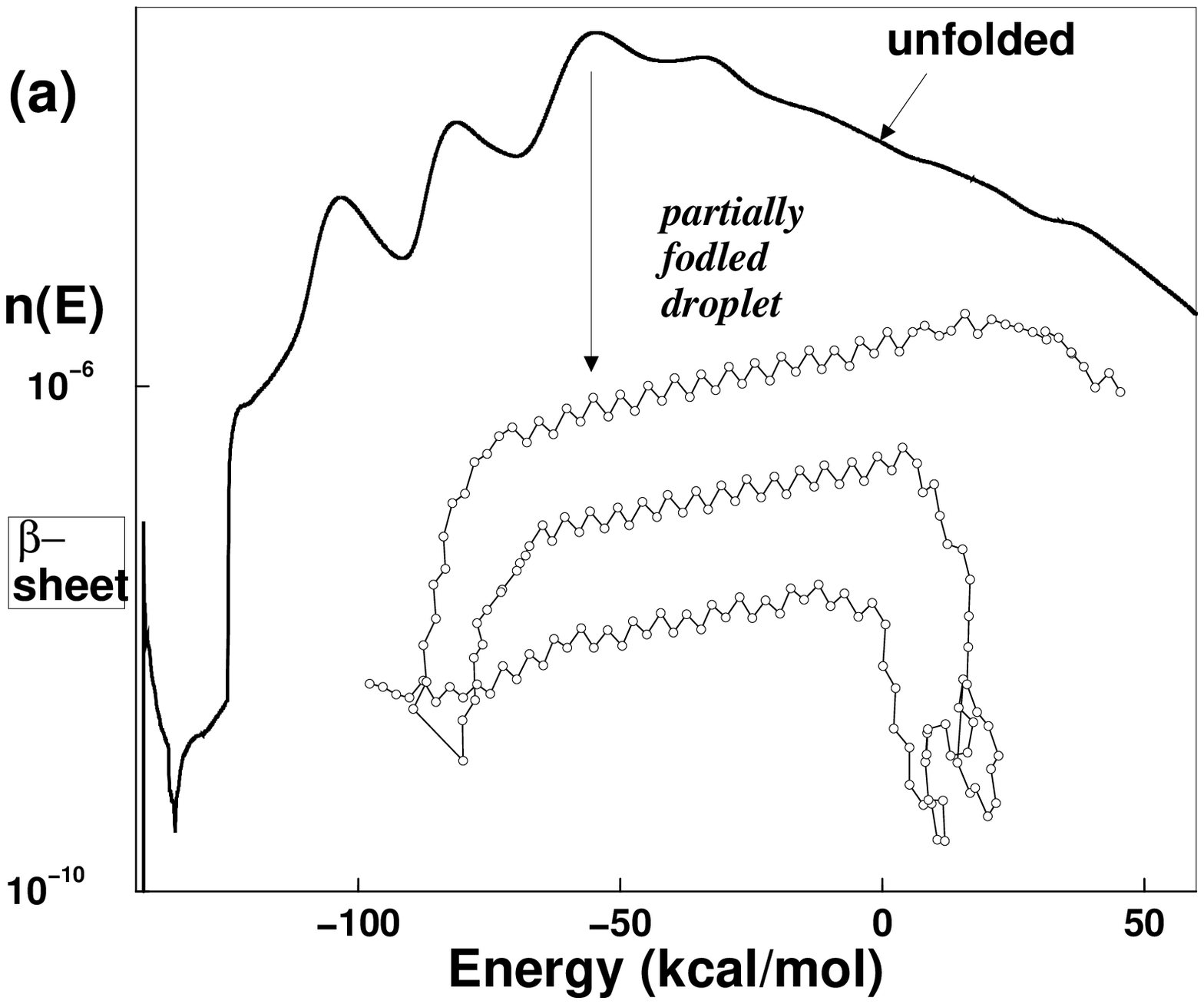}}
{\epsfxsize = 3.2in \epsffile{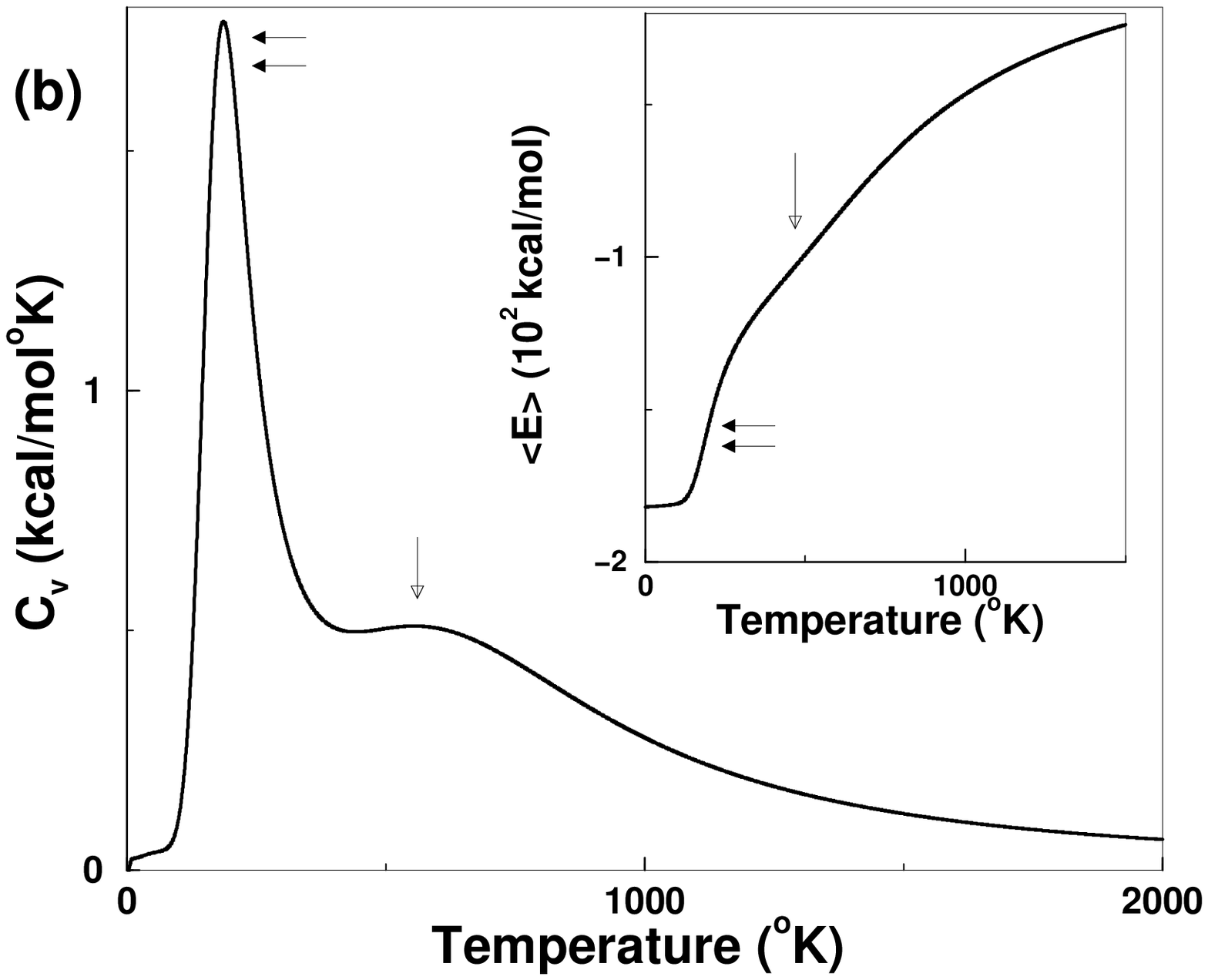}}
}}
\vspace{0.5cm}
\centerline{
{\epsfxsize = 3.2in \epsffile{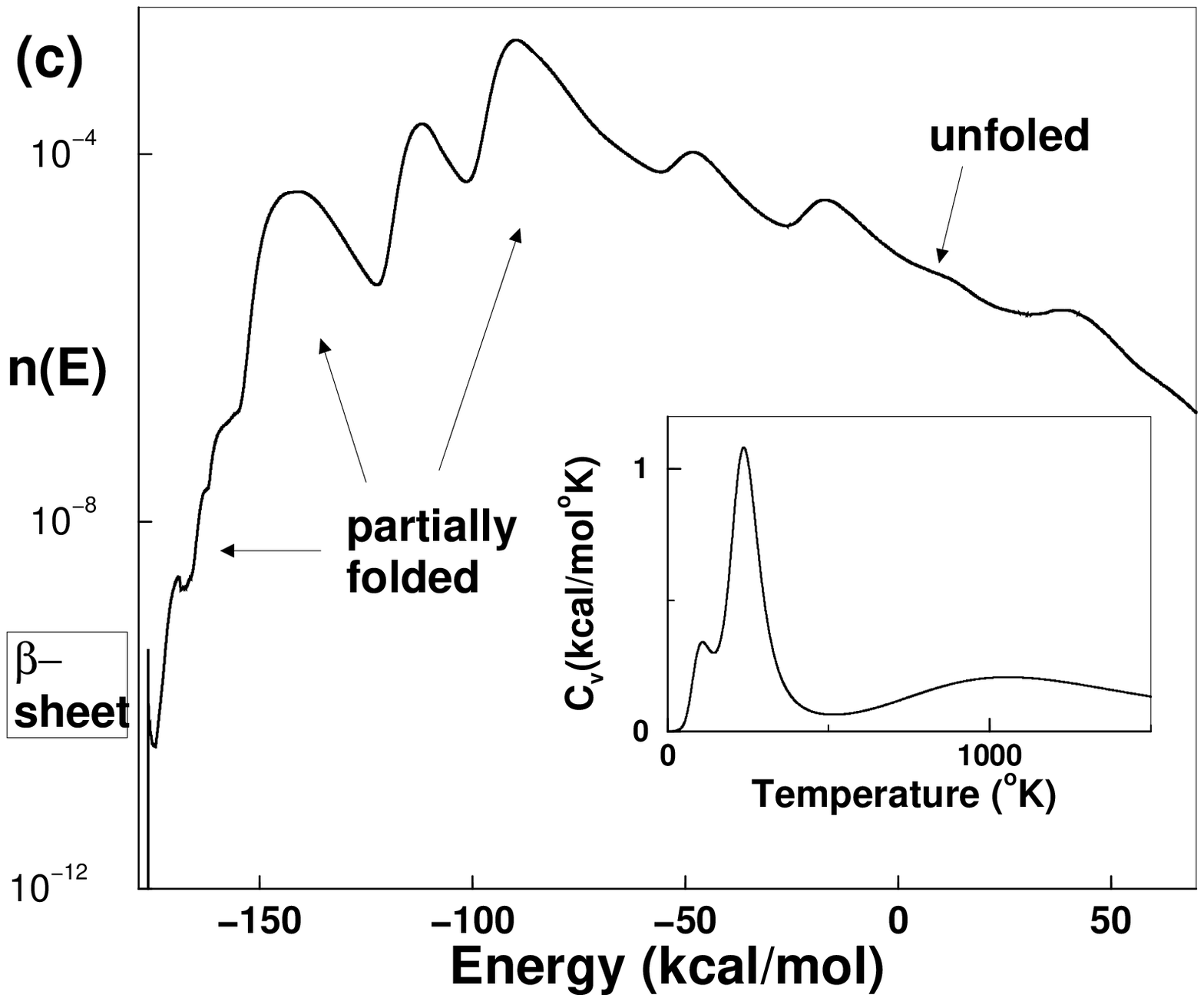}}
}
\vspace{0.5cm}
\caption{}\label{nE4}
\end{figure}

\begin{figure}\centerline{\hbox{
{\epsfxsize = 3.2in \epsffile{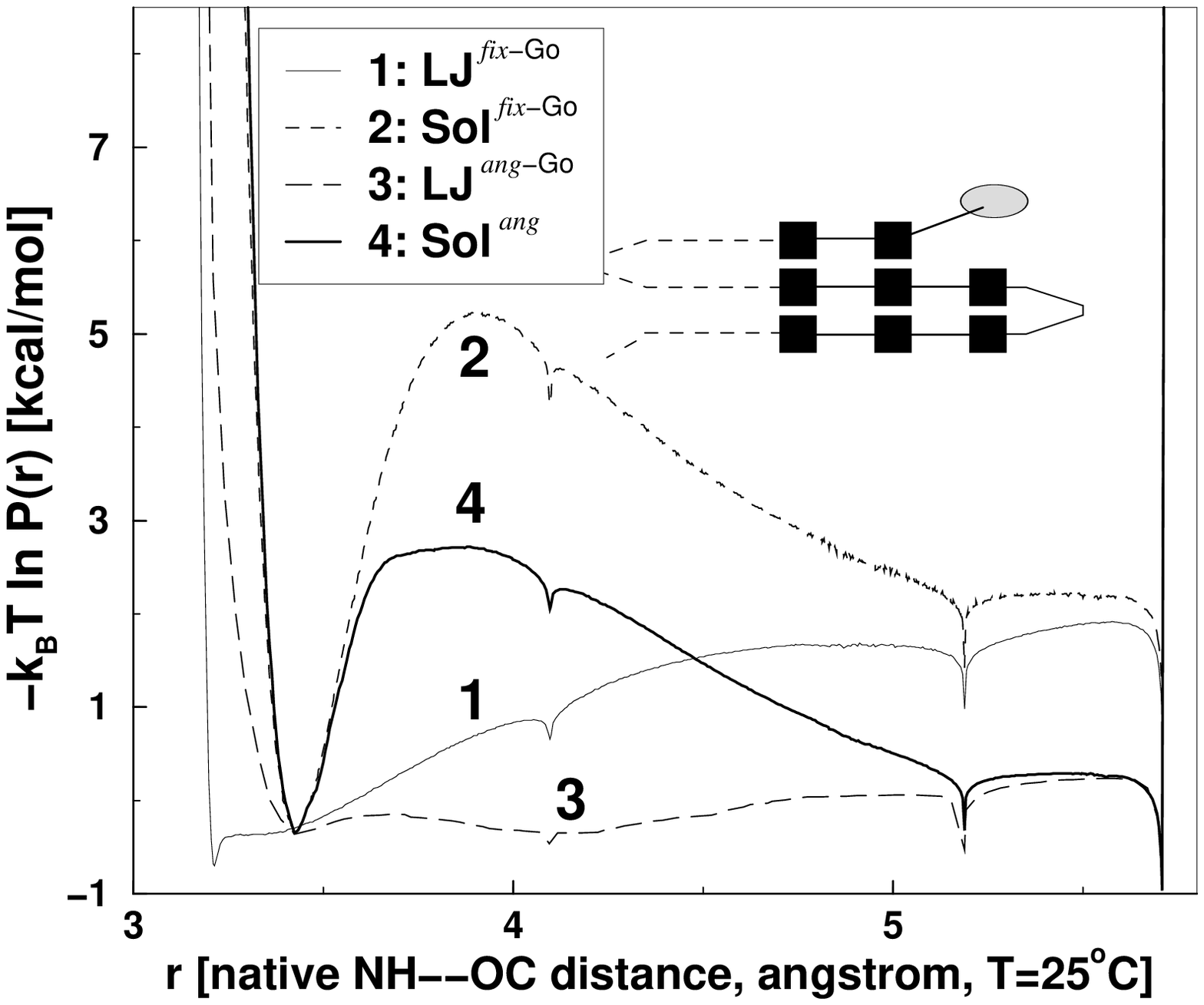}}
}}
\vspace{.5cm}
\caption{}\label{1-block}
\end{figure}

\begin{figure}\centerline{\hbox{
{\epsfxsize = 3.2in \epsffile{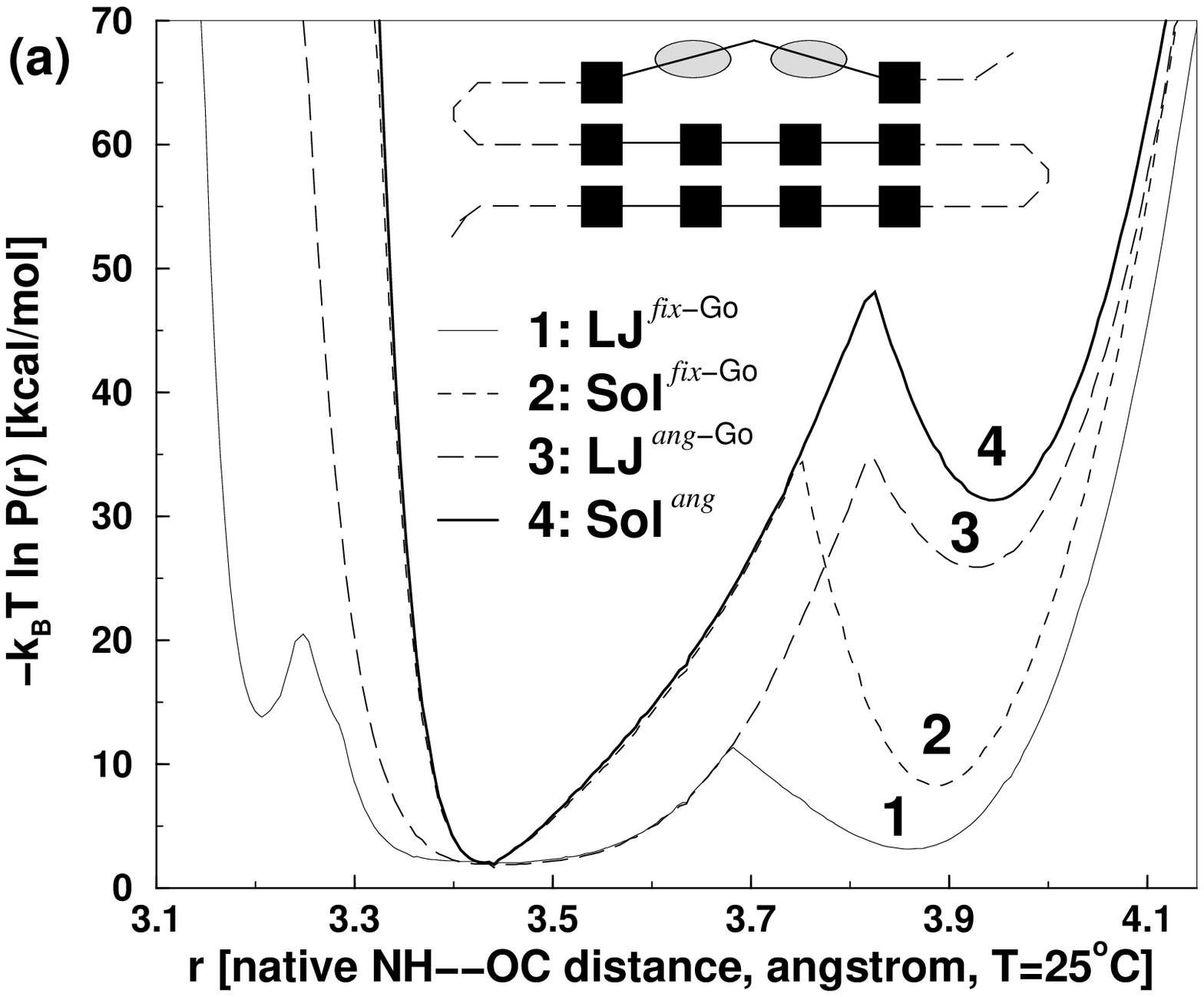}}
{\epsfxsize = 3.2in \epsffile{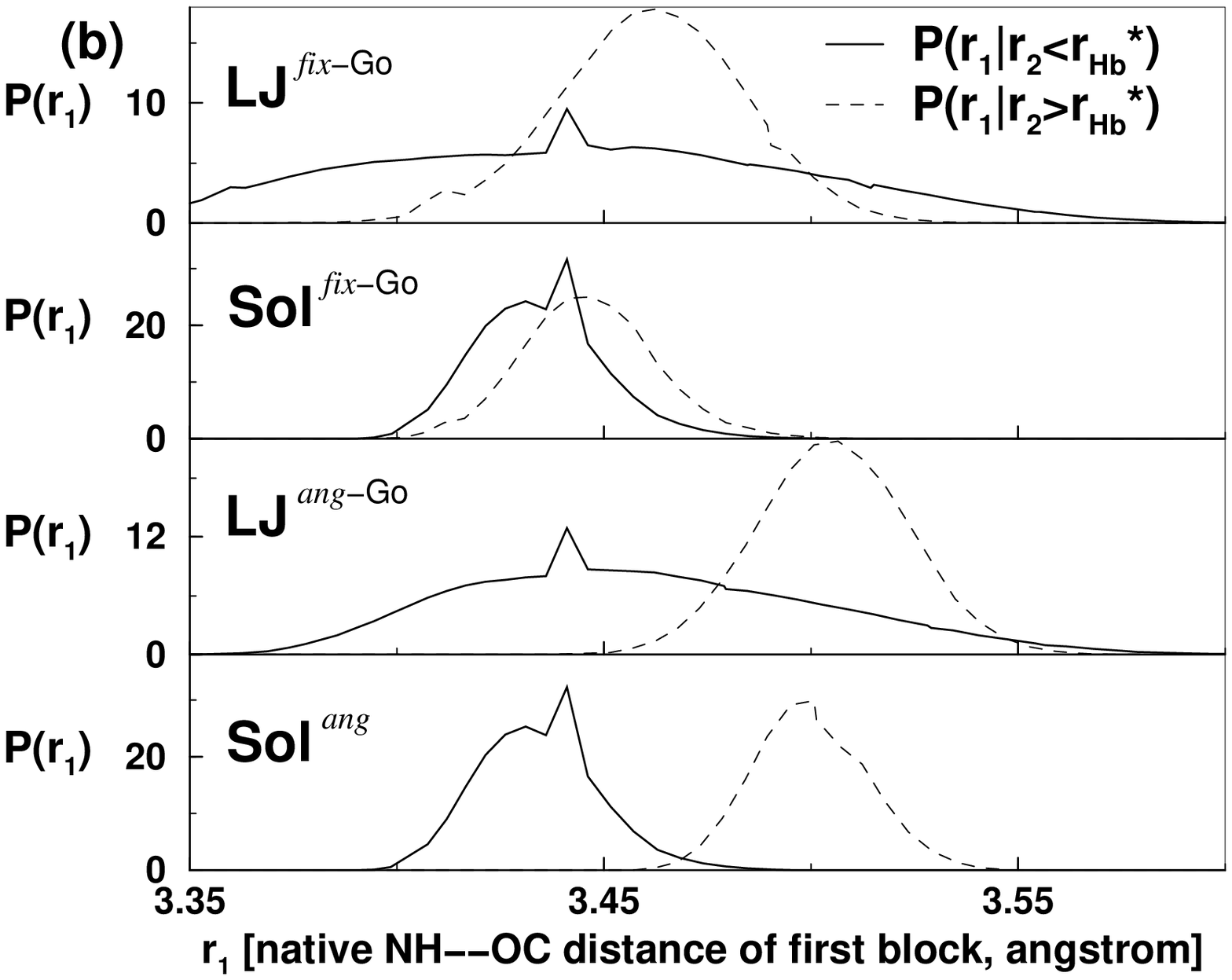}}
}}
\centerline{
{\epsfxsize = 3.2in \epsffile{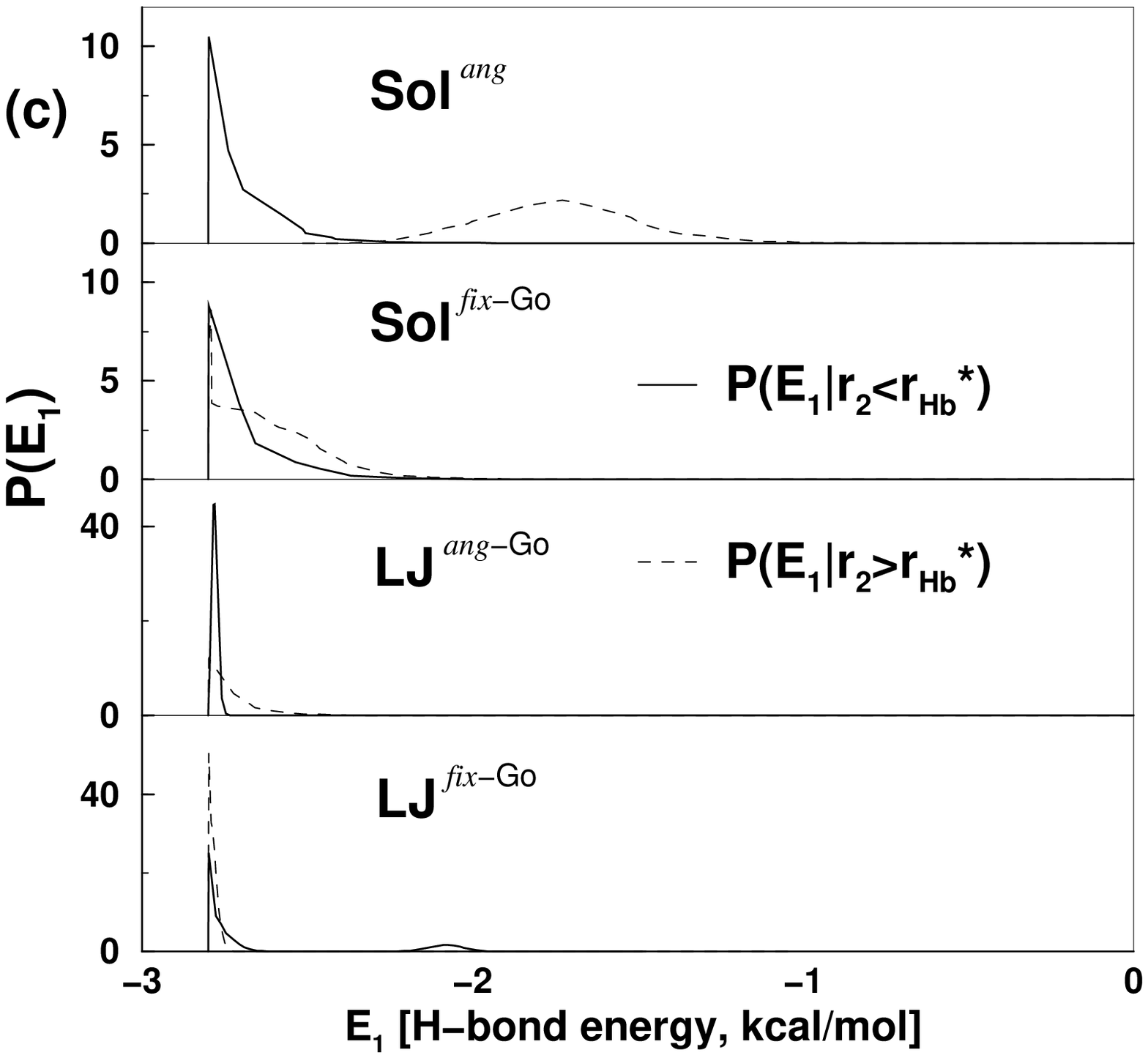}}}
\vspace{.5cm}
\caption{}\label{2-block}
\end{figure}

\begin{figure}
\centerline{
\hbox{
{\epsfxsize = 3.2in \epsffile{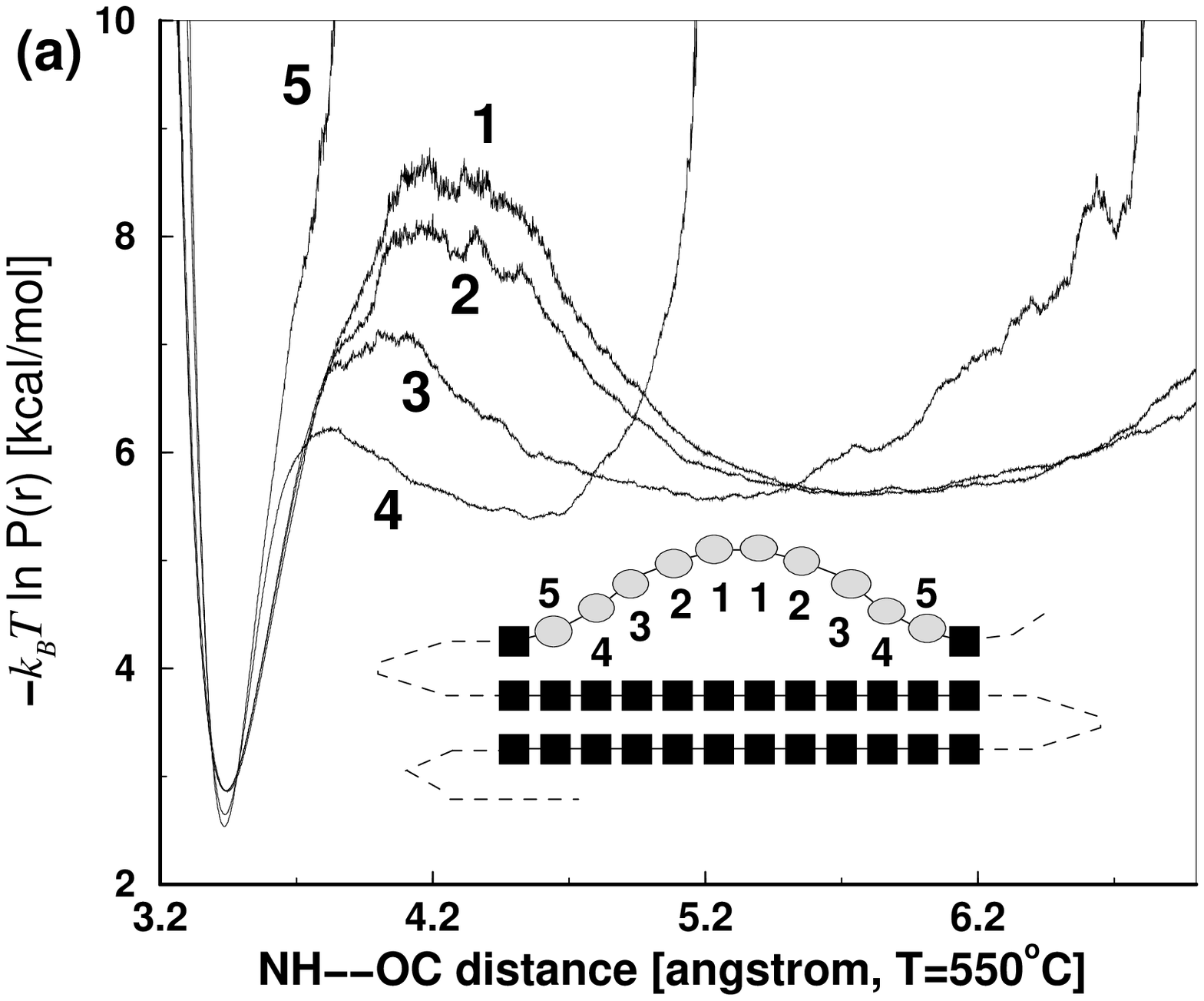}}
{\epsfxsize = 3.15in \epsffile{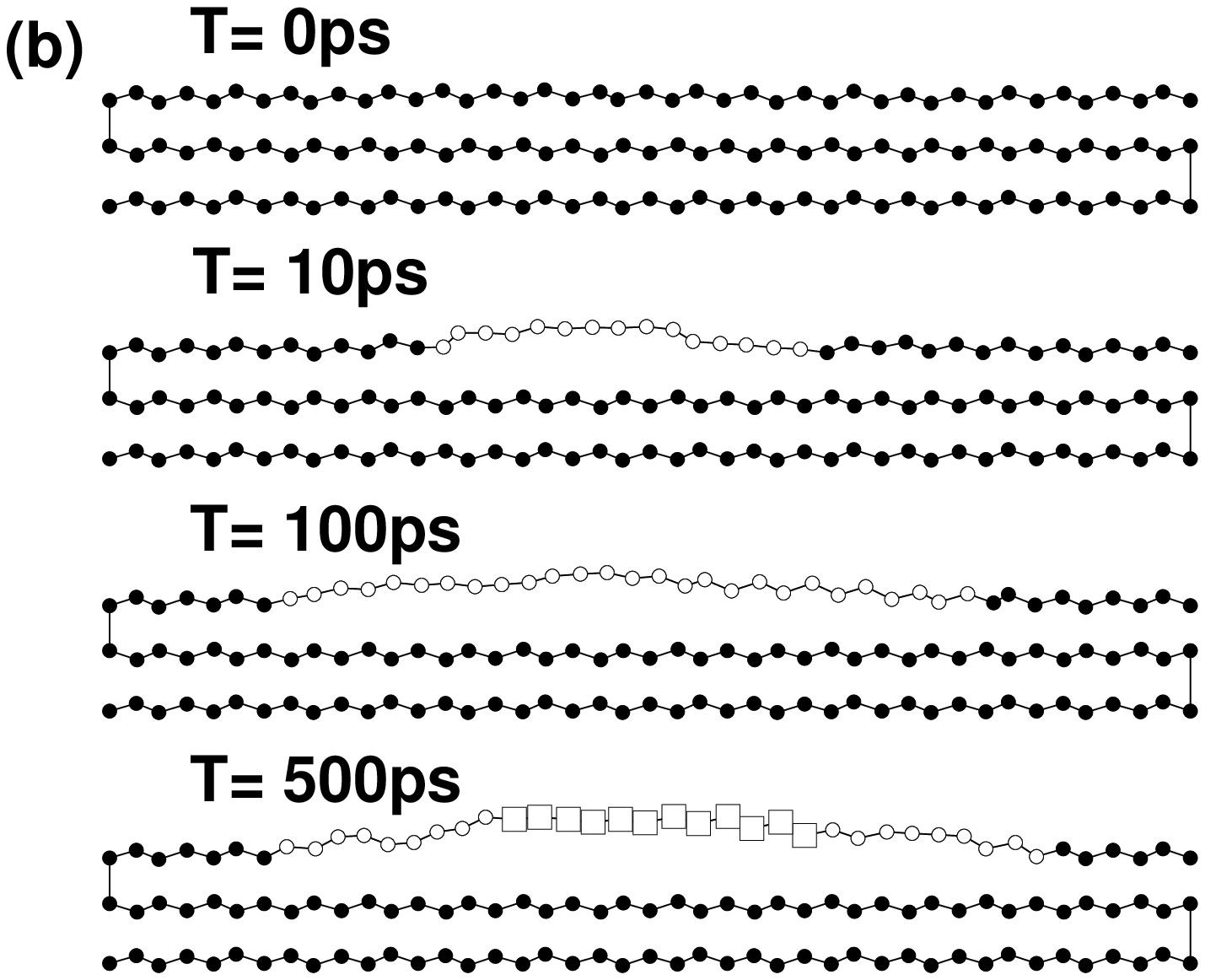}}
}}
\vspace{.5cm}
\centerline{
\hbox{
{\epsfxsize = 3.1in \epsffile{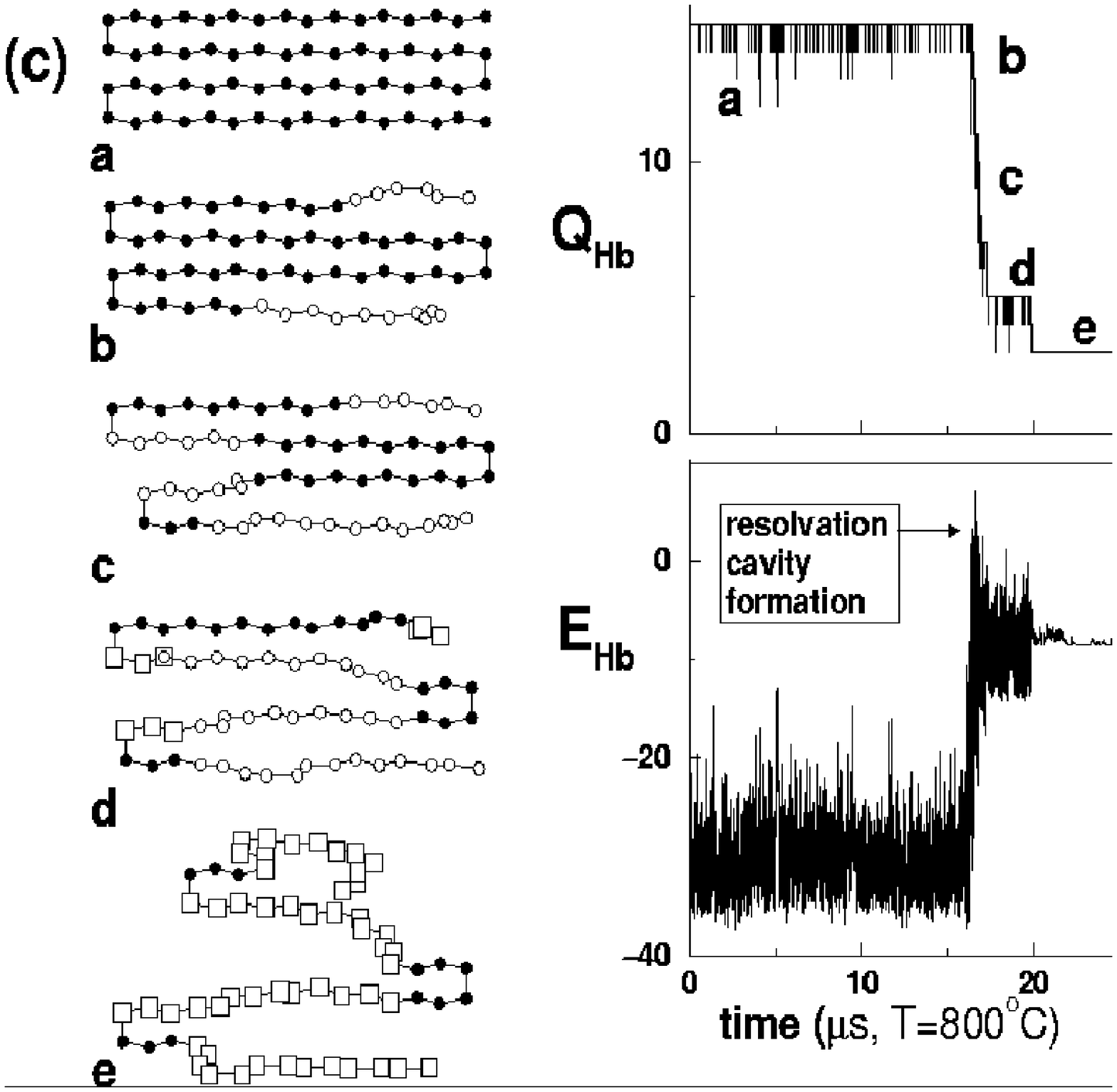}}
}}
\vspace{.5cm} \caption{}\label{kinetics}
\end{figure}

The simplified double-well potential provides the basic insight on how
de/solvation effect involves in protein cooperativity. However, one might
wish to modify this model to include more control parameters. For this
purpose, we propose an alternative potential
\begin{eqnarray*}
{\rm U_{HB}}(r)=
\begin{cases}
-B\frac{(r-r_1)^2-h_1}{(r-r_1)^{2m}+h_2}&\hspace{.5cm}\text{if $r> r_{x1}$,}\\
Ay(r)\left\{[y(r)]^2+b\right\}+{\rm U_{HB}}(r_{x1})
&\hspace{.5cm}\text{if $r\le r_{x1}$}\\
\hspace{0.5cm}\text{with}
\begin{cases}
y(r)=\frac{z(r)[z(r)-1]}{z(r)+s},\\
z(r)=\frac{r_{x1}/r-1}{r_{x1}/r_{x0}-1}
\end{cases}
\end{cases}
\end{eqnarray*}
and $r$ as the distance between NH and CO, fig.\ref{fig0}{\bf(a)}.
Here, the interacting distance for formation of contact,
desolvation barrier, and single-solvent separation are $r=r_0$, $r=r_1$,
$r=r_2$, with free energy amplitude $D_0$, $D_1$, $D_2$, respectively, and
the width of the contact well is controlled by the span between
$r=r_{x0}$ and $r=r_{x1}$. Normally, $r_2$ is separated from $r_0$ by the
size of a single water molecule. The integer $m>2$, on the other hand, is
chosen for the long-distance behavior, and the six unknowns
$B,h_1,h_2,A,b,s$ can be determined by $D_0,D_1,D_2,r_0,r_2$ and one
continuity requirement ($\rm dU_{HB}/dr$ at $r=r_{x1}$).
Using this model, one has a complete control on the potential profile (the
well width for the contact entropy, e.g.).

To incorporate the fully Lennard-Jones behavior into the profile
at both short ($r\ll\rb$) and long ($r\gg\rc$) range region, we
also proposed another force field
\begin{eqnarray*}
{\rm U_{HB}}(r)=A\left(\frac{r_0}{r}\right)^6
\left[\left(\frac{r_0}{r}\right)^6-2\right]
+\frac{Br^{6m}}{1+Cr^{12m}}
\end{eqnarray*}
with $m$ as a positive integer and $A,B,C>0$. Here the second term
is designed to produce the desolvation granularity effect.

\break
\newpage

\end{document}